\documentclass{emulateapj}
\usepackage{amsmath}
\usepackage{multirow}
\usepackage{hyperref}
\usepackage{mathrsfs}
\newcommand{\HST}{{\em HST}}
\newcommand{\AF}{{\sc arcfinder }}
\newcommand{\CF}{{\sc CFHTLS }}
\newcommand{\CW}{{\sc CFHTLS-WIDE }}
\newcommand{\CD}{{\sc CFHTLS-DEEP }}
\newcommand{\WD}{{\sc WIDE }}
\newcommand{\DP}{{\sc DEEP }}
\newcommand{\SA}{{\sc SARCS }}
\newcommand{\vecb}[1]{{\boldsymbol #1}}
\newcommand{\dx}{{\rm d}}
\newdimen\hdsize

\shorttitle{The SARCS sample}
\shortauthors{More et al.}

\begin{document}

\title{The CFHTLS-Strong Lensing Legacy Survey (SL2S): Investigating the
group-scale lenses with the SARCS sample }

\author{A.\ More\altaffilmark{*,1,2}, R.\ Cabanac\altaffilmark{3,4},
  S.\ More\altaffilmark{1}, 
  C.\ Alard\altaffilmark{5}, 
  M.\ Limousin\altaffilmark{2},
  J-P.\  Kneib\altaffilmark{2},
  R.\ Gavazzi\altaffilmark{5} and 
  V.\ Motta\altaffilmark{6}
}
 \altaffiltext{*}{
Based on observations obtained with MegaPrime/MegaCam, a joint
project of CFHT and CEA/DAPNIA, at the Canada-France-Hawaii Telescope (CFHT)
which is operated by the National Research Council (NRC) of Canada, the Institut
National des Science de l'Univers of the Centre National de la Recherche
Scientifique (CNRS) of France, and the University of Hawaii. This work is based
in part on data products produced at TERAPIX and the Canadian Astronomy Data
Centre as part of the Canada-France-Hawaii Telescope Legacy Survey, a
collaborative project of NRC and CNRS.} 
 \altaffiltext{1}{Kavli Institute for Cosmological Physics, U. of Chicago, 5640 S. Ellis Ave., Chicago IL-60637, USA; anupreeta@kicp.uchicago.edu}
 \altaffiltext{2}{Laboratoire d'Astrophysique de Marseille, 38 rue Frederic
 Joliot Curie, 13013 Marseille, France}
 \altaffiltext{3}{Universit\'e de Toulouse, UPS-OMP, IRAP, Tarbes, France }
 \altaffiltext{4}{CNRS, IRAP, 57, Ave. d'Azereix, 65000 Tarbes, France }
 \altaffiltext{5}{Institute d'Astrophysique de Paris, France }
 \altaffiltext{6} {Universidad de Valparaiso, Departamento de Fisica y
 Astronomia, Avenida Gran Bretana 1111, Valparaiso, Chile }

\begin{abstract}
We present the Strong Lensing Legacy Survey - ARCS (SARCS) sample compiled from
the final T0006 data release of the Canada-France-Hawaii Telescope Legacy Survey
(CFHTLS) covering a total non-overlapping area of 159~deg$^2$. We adopt a
semi-automatic method to find gravitational arcs in the survey that makes use of
an arc-finding algorithm. The candidate list is pruned by visual inspection and
ranking to form the final \SA sample. This list also includes some
serendipitously discovered lens candidates which the automated algorithm did not
detect. The \SA sample consists of 127 lens candidates which span arc
radii~$\sim2\arcsec-18\arcsec$ within the unmasked area of $\sim$150~deg$^2$.
Within the sample, 54 systems are promising lenses amongst which, we find 12
giant arcs (length-to-width ratio~$\ge8$). We also find 2 radial arc candidates
in SL2SJ141447+544704. From our sample, we detect a systematic alignment of the
giant arcs with the major axis of the baryonic component of the putative lens in
concordance with previous studies. This alignment is also observed for all arcs
in the sample and does not vary significantly with increasing arc radius. The
mean values of the photometric redshift distributions of lenses corresponding to
the giant arcs and all arcs sample are at $z\sim0.6$. Owing to the large area
and depth of the CFHTLS, we find the largest sample of lenses probing mass
scales that are intermediate to cluster and galaxy lenses for the first time. We
compare the observed image separation distribution (ISD) of our arcs with
theoretical models. A two-component density profile for the lenses which
accounts for both the central galaxy and the dark matter component is required
by the data to explain the observed ISD.  Unfortunately, current levels of
uncertainties and degeneracies accommodate models both with and without
adiabatic contraction.  We also show the effects of changing parameters of the
model that predict the ISD and that a larger lens sample might constrain
relations such as the concentration-mass relation, mass-luminosity relation and
the faint-end slope of the luminosity function.
\end{abstract}


\keywords{dark matter -- gravitational lensing: strong -- methods: data
analysis -- surveys: CFHTLS-SL2S }

%
\section{INTRODUCTION}
\label{sec:intro}

Gravitational lensing is the deflection of light coming from distant sources in
the Universe, due to the gravitational potential of intervening structures
\citep[see reviews e.g.][]{blandford86,blandford92,kochanek06a}. The last decade has seen the rise of a wide variety of
applications of strong lensing such as the study of distant lensed galaxies with
unprecedented magnification
\citep[e.g.,][]{impell08,swinbank09,zitrin09,richard11}, the constraints
on substructure within lensing halos
\citep[e.g.,][]{more09a,suyu10a,vegetti10a,vegetti10b}, accurate
measurements of the Hubble constant \citep[e.g.,][]{coles08,suyu10b},
constraints on the stellar initial mass function
\citep[e.g.,][]{treu10,ferreras10,sonnenfeld11}, constraints on the slope of the inner
density profile of the lensing halos
\citep[e.g.,][]{treu02a,treu02b,koopmans03,koopmans06,more08,barnabe09,koopmans09} and estimation of
the fraction of dark matter in galaxy-scale halos
\citep[e.g.,][]{gavazzi07,jiang07,grillo10,faure11,more11a,ruff11}. 

Although strong lensing is a rare event, several surveys covering a wide sky
area and deep enough imaging across different wavelengths, have resulted in the
discovery of over 200 strong lens systems at galaxy scales from surveys such as, the
Great Observatories Origins Deep Survey \citep{fassnacht04}, Cosmic Evolution
Survey \citep[COSMOS][]{faure08,jackson08},
Mediu Deep Survey \citep{ratnatunga99}
the Cosmic Lens All Sky Survey \citep[CLASS,][]{myers03} and the Sloan Lens ACS Survey
\citep[SLACS,][]{bolton06} and about a few dozen lens systems at cluster scales
such as, the Extended Medium Sensitivity Survey \citep[EMSS,][]{luppino99}, the
MAssive Cluster Survey \citep[MACS,][]{ebeling01}, the Las Campanas Distant
Cluster Survey \citep[LCDCS,][]{zaritsky03} and the Red sequence Cluster Survey
\citep[RCS,][henceforth G03]{gladders03}. Large imaging and spectroscopic
surveys enable us to probe statistical properties of both dark matter and
baryonic matter or constrain cosmological parameters with high accuracy. For
instance, on galaxy scales, the SLACS sample has been used to study the average
density profiles of lens galaxies up to redshift of $\sim$0.3
\citep[e.g.,][]{koopmans06,gavazzi07,koopmans09,auger10}, \citet{falco99} used the CfA-Arizona 
Space Telescope LEns Survey \citep[CASTLES,][]{munoz98} sample
to measure the mean extinction due to the interstellar medium of the lens
galaxies, \citet{mediavilla09} estimated the fraction of mass in compact objects
within lens galaxies from the CASTLES, \citet{wyithe04} constrained the bright
end of the quasar luminosity function from the absence of lensed quasars at high
redshift in the Sloan Digital Sky Survey (SDSS). On cluster scales,
G03 found the observed abundance of giant arcs from the RCS is
too high to be consistent with the predictions from the current standard
cosmological model \citep[see also][]{bartelmann98,li05,li06} and \citet{zitrin11} found
some discrepancy between the predictions from the standard model and the
observed distribution of Einstein radii from the MACS sample. However, the
magnitude of these differences has been mitigated with subsequent studies
\citep[e.g.,][]{horesh05,meneghetti11}. 

As discussed above, the majority of the surveys in the past have primarily
focused on studying galaxy-scale or cluster-scale structures. As a result,
matter distribution in galaxies and galaxy clusters is relatively well-studied
via both strong and weak lensing. A further improvement in our understanding has
come from the use of complementary methods to lensing such as stellar
kinematics, satellite kinematics and X-ray scaling relations. In contrast,
little is known about galaxy groups which are intermediate to galaxies and
galaxy clusters, typically corresponding to masses of
$10^{12}-10^{14}~M_{\odot}$. Relatively fewer investigations have been carried
out with galaxy groups e.g., study of intra-group medium with very low redshift
X-ray sample \citep{helsdon00}, study of mass-to-light ratios with the Canadian
Network for Observational Cosmology 2 sample \citep{parker05}, study of faint
end of the luminosity function of nearby compact groups \citep{krusch06}, study
of concentration-mass ($c$-$M$) relation from the SDSS \citep{mandelbaum08} via
weak lensing, study of colors and star formation
\citep[e.g.,][]{balogh09,balogh11}, study of scaling relations of X-ray selected
groups \citep{rines10} and study of baryon fractions from the 2MASS
\citep{dai10}. Since studies of groups are limited and we still do not have a
detailed understanding of matter distribution, formation and evolution of galaxy
groups. Being one of the important components in the hierarchical assembly of
structures in the Universe, galaxy groups are much more massive than
galaxy-scale halos and are concentrated enough to act as lenses.  Furthermore,
since galaxy groups are quite abundant compared to massive structures like
galaxy clusters, the probability to find group scale lenses is also large.
Hence, lensing can be successfully used to study group-scale halos.

The Strong Lensing Legacy Survey \citep[SL2S,][]{cabanac07} is a survey from the
Canada-France-Hawaii Telescope Legacy Survey (CFHTLS). The design of the CFHTLS
allows us to find large sample of group-scale lenses, which 
can be studied in detail upto high redshifts, for the first time. The SL2S is as a
precursor to wide field imaging surveys such as the Dark Energy Survey, Large
Synoptic Survey Telescope and Euclid. A combined study of SL2S galaxy-scale
lenses with the SLACS sample have been used to show mild evolution of the slope
of the average density profile of galaxies \citep{ruff11} which is constrained
by the strong lensing and stellar dynamics techniques. Disk galaxies are also a
relatively less studied population especially at high redshifts. An automated
search for edge-on disk lenses from the SL2S resulted in 18 candidates , out of
which 3-5 are expected to be real lenses \citep{sygnet10}. A subset of the SL2S
groups have been studied in detail using a combination of techniques such as
strong lensing, group dynamics, weak lensing to probe the density profiles of
the lensing groups \citep[e.g.,][]{limousin09,thanjavur10,verdugo11}.

In this paper, we present the SL2S-ARCS sample from the final T0006 release of
the CFHTLS. In Section 2, we give an overview of the survey and procedure of
sample selection, describe details of the algorithm, {\sc arcfinder}
\citep{alard06} and present the final sample. In Section 3, we discuss some
statistical results using the final sample. In Section 4, we summarize the
survey and our main findings.

\section{THE CFHTLS-SL2S ARCS SAMPLE}
In this section, we give a brief overview of the survey from which we derive the
lens sample. This is followed by a description of the semi-automatic process of
selecting the candidates in the final sample. We also discuss how the algorithm,
\AF works and the modifications implemented in the new {\sc arcfinder}. Lastly,
we present the final sample and report duplicate detections of some candidates
from other surveys.

\subsection{Survey Overview}
\label{sec:ovrw}

CFHTLS is a photometric survey made with the Canada-France-Hawaii Telescope
(CFHT) in five optical bands ($u^*g'r'i'z'$) using the wide-field imager MegaPrime
with a field-of-view of 1~deg$^2$ on the sky and a pixel size of 0.186\arcsec. The \WD
and \DP components of the CFHTLS are designed to carry out extragalactic
research. These components are ideal for searching strong lens systems. The
SL2S sample is compiled from the \CW encompassing a combined area of 171~deg$^2$
and \CD encompassing a combined area of 4~deg$^2$.  However, taking into account
the masked and overlapping areas, the effective area of the survey is
150.4~deg$^2$ (146.9~deg$^2$ for \WD and 3.5~deg$^2$ for DEEP). The \WD consists
of four fields W1, W2, W3 and W4. The field W1 has the largest sky coverage of
63.65~deg$^2$. The fields W2 and W4 have similar sky coverages of 20.32~deg$^2$
and 20.02~deg$^2$, respectively\footnote{These numbers are estimated from
\href{http://terapix.iap.fr/cplt/table\_syn\_T0006.html}{http://terapix.iap.fr/cplt/table\_syn\_T0006.html}}. The field W3 has a sky
coverage of 42.87~deg$^2$ and is more than twice as large as W2 and W4. The \DP
also consists of four fields D1 (located within the W1 field), D2, D3 and D4.
Each of the deep fields covers an area of 1~deg$^2$.  The DEEP images are
produced in two image stacks D-25 and D-85. The former consists of 25\% of the
best seeing images and the latter consists of 85\% of best seeing images. We use
the D-25 images for searching lens candidates. Among the \WD fields, $g$ band
imaging is the deepest of all bands with a limiting magnitude of 25.47 and a
mean seeing of 0.78\arcsec\ whereas the $g$ band imaging of \DP fields has nearly 10
times deeper exposures than the \WD fields and the median seeing is $\sim0.7\arcsec$.
The zero point to convert flux to AB magnitude for all bands is 30. Further
details of the T0006 release, which is the first complete release of the \WD and
DEEP, can be found on Terapix
website\footnote{\href{http://terapix.iap.fr/cplt/T0006-doc.pdf}{http://terapix.iap.fr/cplt/T0006-doc.pdf}}.

\subsection{Sample selection}
\label{sec:sample}

The SL2S lens sample is compiled using two algorithms: {\sc ringfinder} and 
{\sc arcfinder}. The former aims at detecting galaxy-scale lenses by using color
information. The SL2S RINGS sample will be presented and discussed in a separate
paper (Gavazzi et al., in preparation). Here, we focus on the SL2S-ARCS (SARCS)
sample created with the help of {\sc arcfinder}, followed by visual inspection
and screening of the candidates. 

We define the \SA sample such that lens systems with arc radius ($R_{\rm A}$) 
$\gtrsim2\arcsec$ belong to the sample. The radius of the arc is defined
as the distance of the lensed image from the putative lens galaxy which is
roughly the Einstein radius. Typically, lensing halos with Einstein radius
larger than 2\arcsec are very massive lenses with significant contribution from
the environment of the primary lensing galaxy. Thus, the \SA sample,
predominantly, consists of group to cluster scale lenses. Lens systems with
$R_{\rm A} < 2\arcsec$, typically, form part of the {\textsc RINGS} sample. We note
that a few lens systems are common to both SL2S samples since the cut on
$R_{\rm A}$ is not a sharp limit imposed by the algorithms.

The \SA sample from the CFHTLS is compiled in a three-step process. The first
step is to run the arc-detection algorithm called {\sc arcfinder}. We choose to
run the \AF on the $g$-band image since most of the lensed images correspond to
galaxies with high star formation that have little emission at redder
wavelengths. Also, focusing on g-band prevents from detecting high redshift
g-dropouts. However, we plan to search for g-dropout arcs which are
brighter in the $i$ band. These results will be presented in a separate paper.

At the end of the first step, we produce a list of arc candidates with various
parameters. The next step is applying a cut-off on arc properties such as the
area, the peak flux count, the radius of curvature and reject candidates within
masked area.  These cuts allow a significant reduction in false detections at
the cost of losing some real arc candidates. In the third step, visual
inspection and classification are carried out to grade the quality of the
candidates. Note that the final \SA sample consists of candidates that are
detected by the \AF and/or by visual inspection. 

\subsection{Automatic detection of arcs: \AF} 
\label{sec:afin}

With the advent of several large imaging surveys, searching for lens systems
visually is a subjective and time expensive exercise which may not be easily
repeatable. Hence, several algorithms have been devised in the last few years to
automate the process of lens detection as much as possible. The algorithms,
usually, focus on a specific target population, for example, algorithms to
detect the lensed sources such as quasars via the quasar time variability
\citep{kochanek06} based on the difference imaging technique of \citet{alard98},
spectrum-based algorithms such as the one used by \citet{bolton04} to form the
SLACS sample following the technique of \citet{warren96} or lens-modelling
robots that assesses the probability of a bright red galaxy being a lens
\citep{marshall09}. Since groups to cluster scale lenses form arc-like lensed
images, there are a few arc-finders in the literature
\citep[e.g.,][]{lenzen04,horesh05,seidel07} that aim at detecting elongated
arc-like images. 

The \AF \citep{alard06} is a generic algorithm that can be used to detect
elongated and curved features in an image. The algorithm uses pixel intensities
from a standard FITS image to trace the structure of a feature. Multiple
thresholds are applied to the structural properties of the feature to select an
arc candidate. The reader is referred to \cite{alard06} for the details of the
algorithm. Below, we describe the algorithm along with some modifications
implemented in the newer version (V2.0) which is used to compile the SARCS sample. 

\begin{figure}
\begin{center}
\includegraphics[scale=0.27]{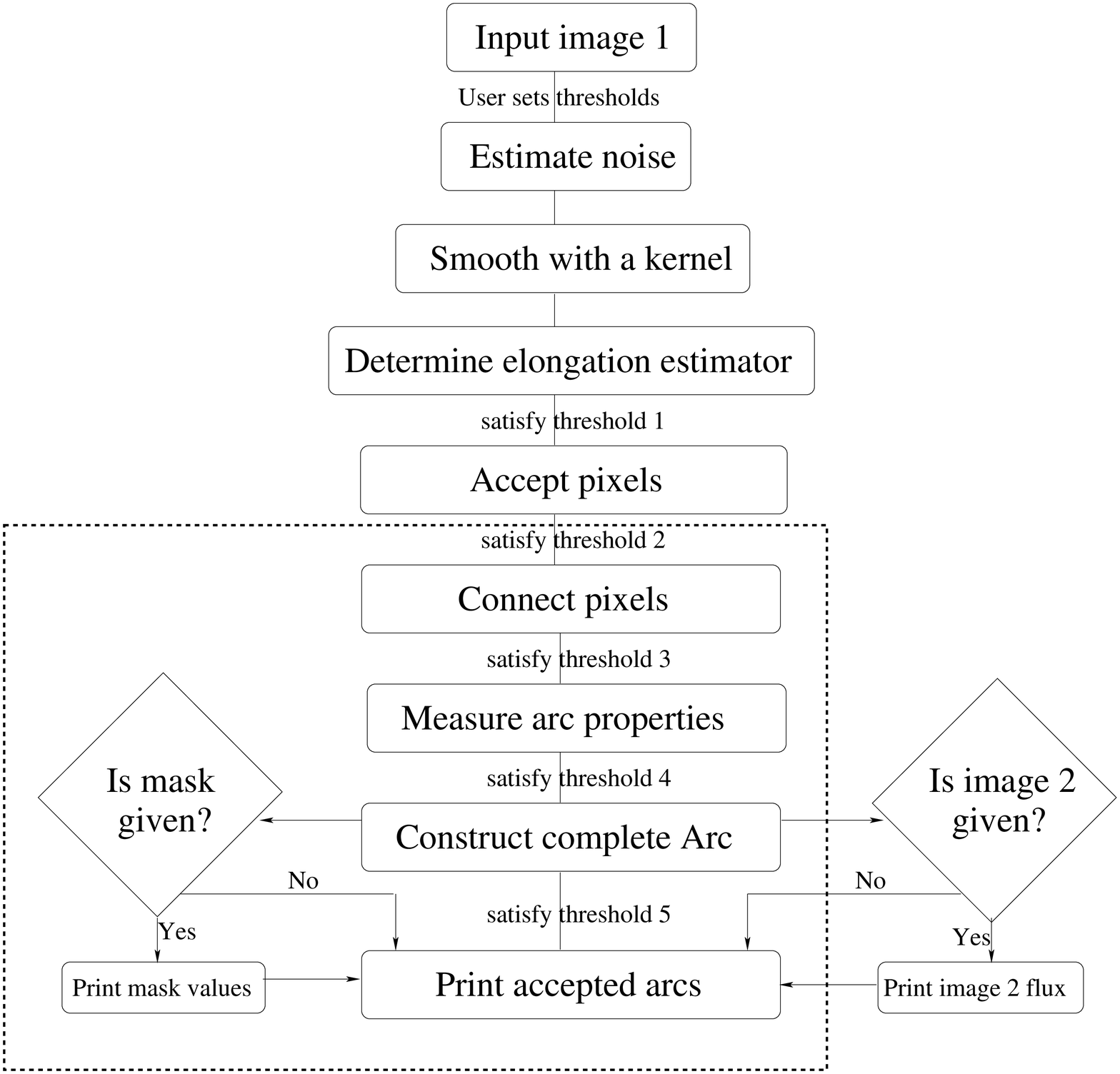}
\caption{ \label{fig:algo}
Flowchart showing the {\sc arcfinder} algorithm.}
\end{center}
\end{figure}

\begin{figure*}
\begin{center}
\includegraphics[scale=0.25]{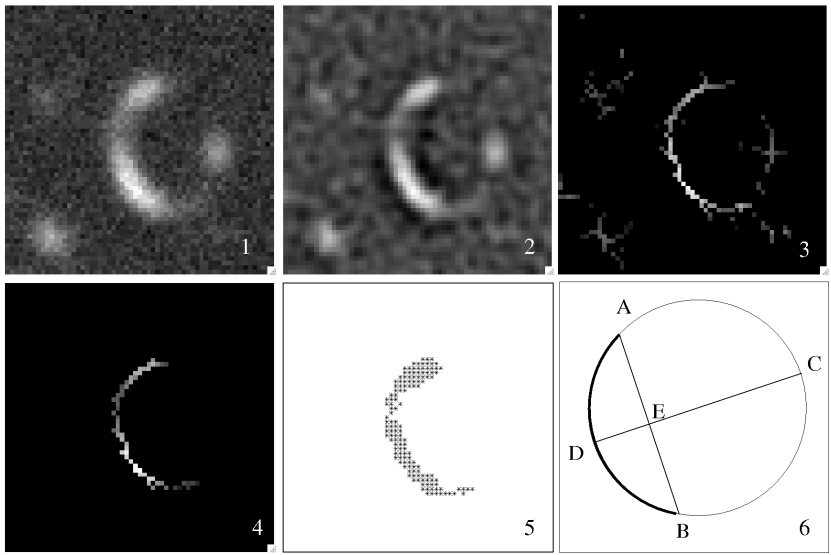}
\caption{ \label{fig:stp}
Outputs of \AF at various stages of the execution (panels 1-5) using a mock arc
image. Panel 1 - original image with the mock arc, panel 2 - smoothed image,
panel 3 - (disconnected) pixels with elongation estimator values higher than a
threshold, panel 4 - connected pixels above thresholds that belong to the arc
and panel 5 - the reconstructed shape of the final arc. The image is 11.3\arcsec on
the side. Schematic diagram for measuring length of an arc (panel 6).
}
\end{center}
\end{figure*}

We show a flowchart to illustrate the various steps involved in the algorithm
(see Fig.~\ref{fig:algo}) and step through an example of a mock arc image (see
Fig.~\ref{fig:stp}). 

\begin{itemize}
\item We run the \AF on images of 19354$\times$19354 pixels which
corresponds to a single MEGACAM pointing. For such large images, the assumption
of Gaussian distribution for the noise holds well. Assuming a Gaussian
distribution for the noise in the input image (panel 1 of Fig.~\ref{fig:stp}),
the noise $\sigma$ is calculated in the first step. Most of the arcs are
unresolved with the ground-based telescopes like CFHT and are limited by the
size of the point spread function (PSF). 

\item Convolving the image with a smoothing
kernel, with a size of the order of the PSF, can help damp the noise and enhance
the detection of arcs. Hence, the input image is smoothed with a Mexican-hat
filter in the next step (panel 2 of Fig.~\ref{fig:stp}). 

The Mexican-hat band-pass filter with scale $b$ is given by
\begin{equation}
 M(x,y)=e^{-\bar r^2}-0.5e^{-\frac{\bar r^2}{2}} \, ,
\end{equation}
where $b^2=3$~pixels and
\begin{equation}
 \bar r=\frac{\sqrt{x^2+y^2}}{b} \, .
\end{equation}


\item Now, we define a local estimator of elongation for every pixel in the smoothed image.
This is calculated by using the flux from all pixels within a window centered on
every pixel. The size of the window is chosen to be a few times the PSF size.
We consider a square window of side W centered on the pixel at (x,y). Using the second
order moments of brightness distribution within the window, the local direction of
the elongation of the feature is calculated. This direction is then used
to align the feature along the x-axis to determine the local axis ratio, similar
to a length-to-width ratio, within the window. The elongation estimator is
defined as follows

\begin{equation} 
Q(x,y)= \frac{1}{W} \frac{F_{\rm X}}{F_{\rm Y_{\rm max}}} \, ,
\end{equation}

where $F_{\rm X}$ is the integrated flux of the single central row of pixels and
$F_{\rm Y_{\rm max}}$ is the maximum of the integrated flux values calculated
from single columns of pixels within the window. If $F_{\rm X}$ and $F_{\rm
Y_{\rm max}}$ are above $(W+1)0.75\sigma$ (threshold 1 in Fig.~\ref{fig:algo}
and Table~\ref{tab:thresh}), then the estimator $Q(x,y)$ is assigned to the
pixel (x,y) at the center of the window else the pixel is assigned a value of 0.
Likewise, the estimator is calculated for every pixel in the image (see panel 3
of Fig.~\ref{fig:stp}). 


\begin{table}
\begin{center}
\caption{ \label{tab:thresh} 
Thresholds used in the final selection of the \SA candidates. }
\begin{tabular}{ll}
\tableline
\tableline
Step & Parameters and Thresholds  \\
\tableline
Smooth & W=9, $b^2=3.0$  \\
Detect & W=11, Th1=11*0.75$\sigma$ \\
Connect & W=9, Th2=1.25 \\
 & W=9, Th4=0.5S1,0.3S2 \\
\\
\tableline
\tableline
Properties & Final Thresholds \\
\tableline
Area & $55<~{\rm A}~<500$~(pix)  \\ 
peak counts & $<50$~(ADU~s$^{-1}$)   \\
mean counts & $2\sigma<~{\rm SB}~<50$~(ADU~s$^{-1}$~pix$^{-1}$) \\ 
length  & $>7$~(pix)   \\
width & $1.5\le~{\rm w}~\le8$~(pix) \\ 
curvature & $r_c<100$ and $r_c>1000$~(pix)  \\
\tableline
\end{tabular}
\tablecomments{W is the window size in pixels. Th1, Th2 and Th4 imply thresholds
1, 2 and 4, respectively. S1 and S2 imply intermediate mean counts of candidates
in the input and smoothed images, respectively.}
\end{center}
\end{table}

\item In the following step, we attempt to connect the pixels that possibly
belong to the candidate arc. A pixel is accepted if the estimator value from
the image in panel 3 is $>1.25$ (threshold 2 in Fig.~\ref{fig:algo} and
Table~\ref{tab:thresh}). The estimator is set to 0 as soon as a pixel is
connected in the prior step to avoid repetitive checking of connected pixels.
Iterating the above steps over the image allows tracing of the primary shape of
the arc. If the number of connected pixels is more than 10 (threshold 3) then
the properties of arcs such as the length, peak flux and surface brightness are
calculated. The panel 4 shows the image with connected pixels which, for the
case at hand, consists of one candidate only. The estimator is particularly
suited for recovering pixels along the length of the arc. In order to connect
the pixels along the width of the arc, we use the input and smoothed images
(panels 1 and 2, respectively). Using arcs with length $>7$ pixels, we apply surface brightness
thresholds on pixel values from images in panels 1 and 2 in order to accept
pixels belonging to the arc (threshold 4, see Fig.~\ref{fig:algo} and
Table~\ref{tab:thresh}) and construct the arc fully. 

\item In the last step, if the candidate arc satisfies thresholds on the arc
properties such as the width ($\ge1.5$ and $\le$8 pixels), area ($>25$ pixels),
surface brightness ($>2\sigma$) and peak flux ($<500$ ADU/s/pixel) which
correspond to threshold 5 in Fig.~\ref{fig:algo}, then the arc is accepted (see
panel 5 in Fig.~\ref{fig:stp}).

\item In the following, we describe how the arc properties are measured. The area of
the arc is the total number of pixels belonging to the final arc. The length of
the arc is calculated by assuming that the candidate arc is similar to an
arc of a circle. We first find the extreme ends of the arc, connect them with a
chord labeled AB in the panel 6 of Fig.~\ref{fig:stp}. From the midpoint E of
chord AB, we draw a line perpendicular to the chord and find its intersection D
with the arc. The lengths ED and AB uniquely identify the circle to which the arc
belongs and as well as the length of the arc. The width of the arc is
defined as the ratio of the area to the length of the arc. The radius of
the circle going through the arc is used as a proxy for the curvature of the
arc.
\end{itemize}

In the algorithm, the use of a second image at another wavelength and a mask
file are optional. The algorithm, in its current version, merely prints the
values of the pixels belonging to the candidate arc from the image 2 and/or the
mask file, if provided. This allows us to optionally screen the candidate arcs
based on their color and/or masking information. For the \SA sample, we made use
of the mask option only and further restricted the parameter space of the output
list of arcs by introducing more strict cuts on some of the arc properties.
The Table~\ref{tab:thresh} lists the final set of thresholds which
are satisfied by the \SA candidates detected by {\sc arcfinder}.

The components enclosed by the dashed line in the Fig.~\ref{fig:algo} are the
sections of the algorithm that have been improved. This includes
the way in which the pixels belonging to the arc are connected and the way the
properties of the arcs are measured. In the earlier version of the algorithm, a
candidate arc is accepted at the final stage only if it satisfies a certain
threshold on the curvature of the arc. This feature has been removed and the
option of using mask information is introduced in the V2.0. The values
for all the different thresholds are tuned from an initial sample of \CF arcs,
found visually or from the previous version of {\sc arcfinder}, based on an early
release of the data. Furthermore, all the thresholds can now be set during
the execution of the algorithm.

The \AF V2.0 finds $\sim 1.5 \times$ more detections, most of which arise
due to spikes and halos near stars. However, the use of masks removes these
false detections thereby making the number of detections comparable or less than
the output of the earlier version. The modified \AF is over 3 times faster
than the earlier version. We have also recently parallelized the algorithm which
enables us to achieve even faster computation on shared memory platforms.
The \AF V2.0 is available upon request to the author.

Finally, we note that the existing algorithms are far from perfect and almost
always require manual intervention. Attempts to increase the completeness of the
sample almost always leads to a corresponding increase in the rate of false
positives. In light of these issues, citizen science projects such as the Galaxy
Zoo \citep{lintott08} might be a better tool, for the time being, in identifying
lens systems not detected by the algorithms. These projects, in turn, could be
used to calibrate and improve existing algorithms. We are currently pursuing the
feasibility of such a project. 

\subsection{\SA sample}

\begin{figure*}
\begin{center}
\includegraphics[scale=0.65]{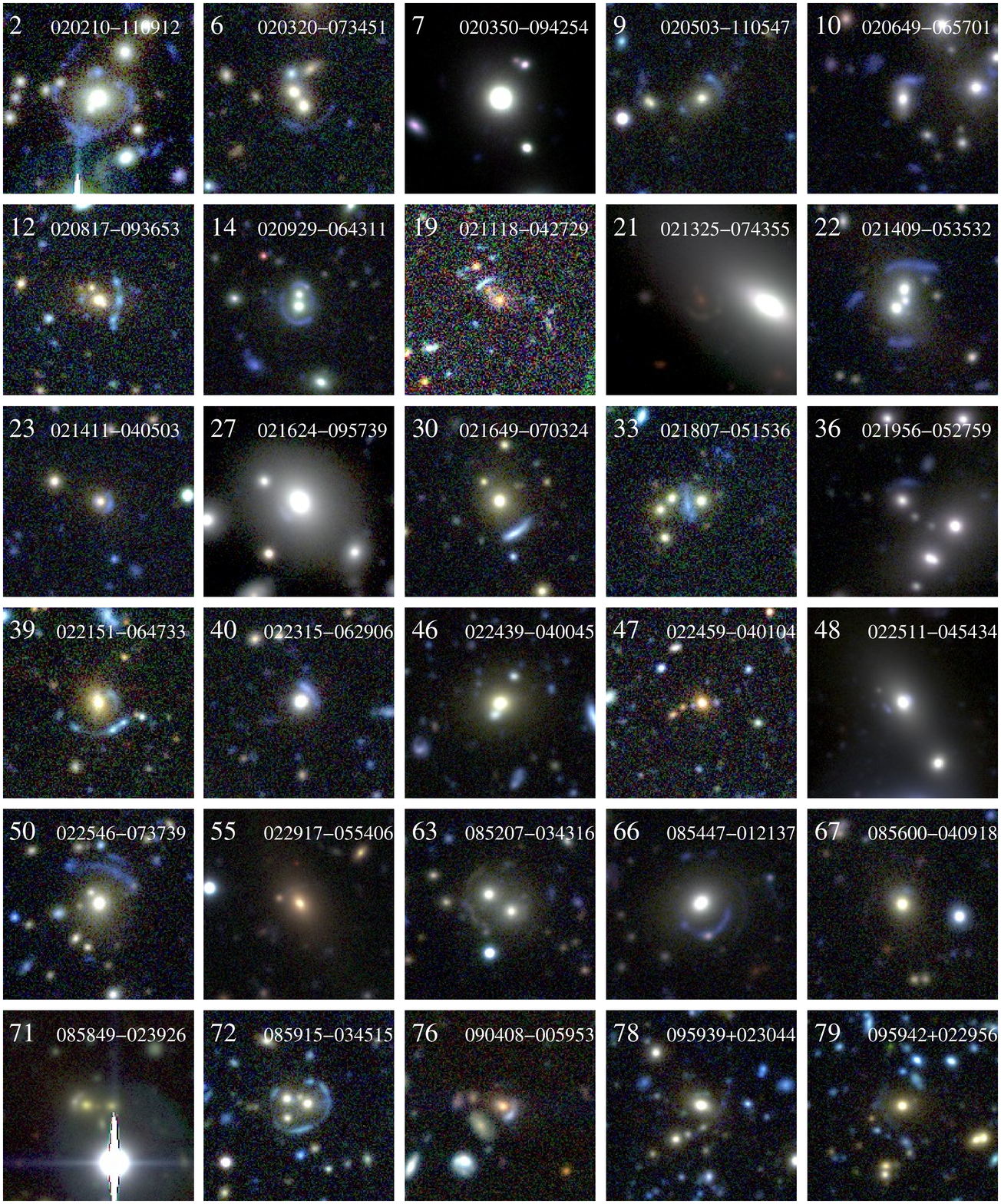}
\caption{ \label{fig:coll1}
The \SA sample showing the 54 promising lens systems with rank of 3 and above.
All color cutouts are made from CFHT imaging in $g,r$ and $i$ bands. The
cutouts are $\sim$30\arcsec on the side. The candidate SA78 is known as COSMOS
 5939+3044.
}
\end{center}
\end{figure*}

\begin{figure*}
\begin{center}
\includegraphics[scale=0.65]{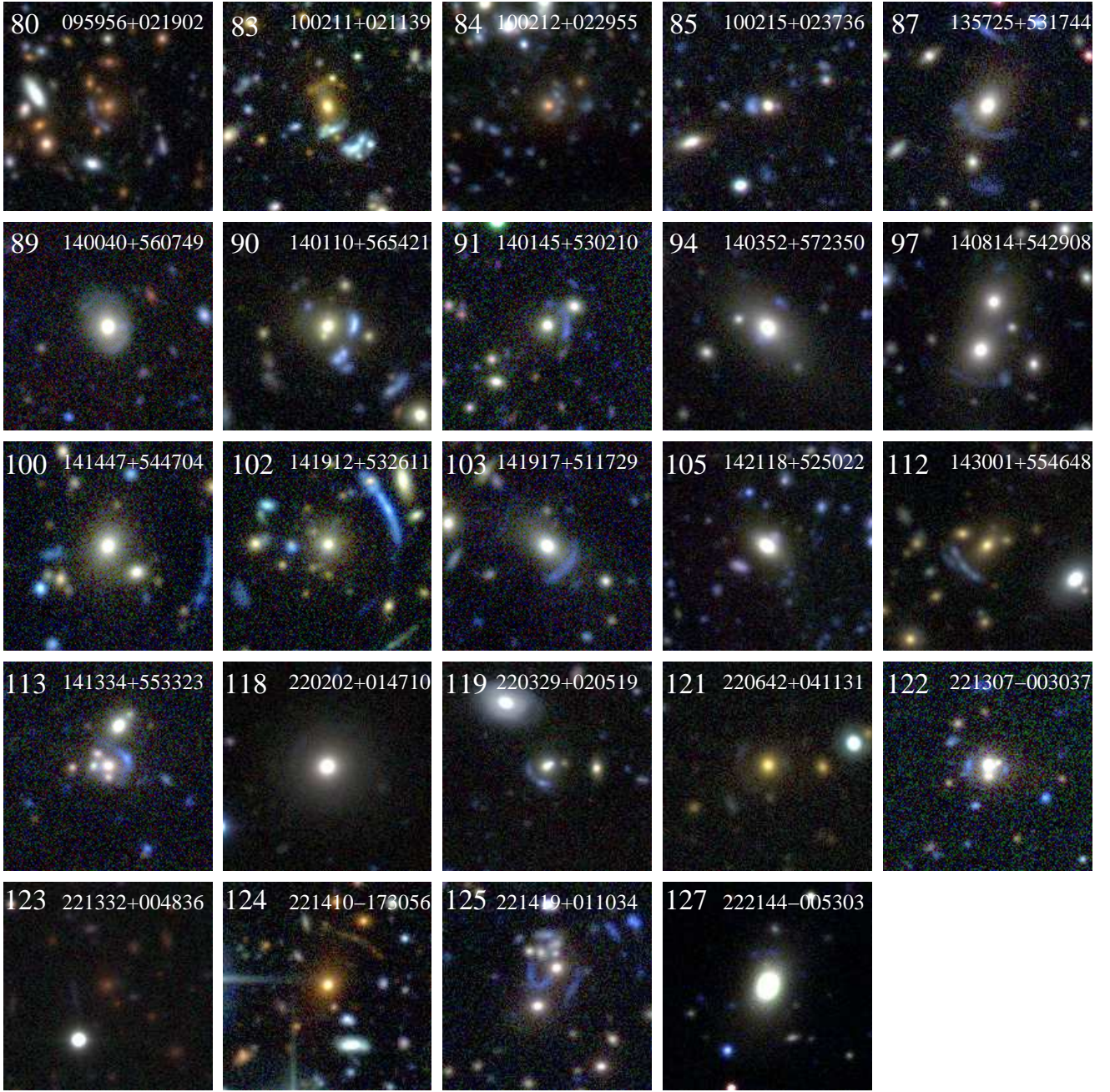}
\caption{ \label{fig:coll2}
Continue Fig.~\ref{fig:coll1}. The candidates SA83 and SA102 are known as COSMOS
0211+1139 and RCS 1419.2+5326, respectively.}
\end{center}
\end{figure*}

After the automatic detection and screening, about 1000 candidates/deg$^2$ are
visually inspected. This is reduced to a total sample of 413 candidates which is
further considered for ranking by three people. The individually assigned ranks
are from 1 to 4 with 1 being the least likely to 4 being the almost certain lens
system. We present the 127 lens candidates which are ranked 2 or higher (average
of the ranks by three people) and which have $R_{\rm A} \gtrsim 2\arcsec$. All
of the 127 candidates \footnote{ High resolution images of these systems are
made available at
\href{http://kicp.uchicago.edu/$\sim$anupreeta/sarcs\_sample}{http://kicp.uchicago.edu/$\sim$anupreeta/sarcs\_sample}.}
are listed in Table~\ref{tab:all} that gives the ID, lens position, lens magnitudes in
AB, lens redshifts with 1$\sigma$ uncertainties, arc radii, ranks, the type indicating whether the candidates are
primarily detected via the {\sc arcfinder} and the field name in which the
candidate is located. For the calculation of arc radii in physical units, we use
flat cosmology with following parameters
($\Omega_m,\Omega_\Lambda,H_0$)=(0.3,0.7,100~km~s$^{-1}$~Mpc$^{-1}$).  The
symbol A stands for detection by \AF and V implies candidate is found
serendipitously.  A total of 54 candidates are good to best systems (that is,
rank of 3 and above) and are shown in the Fig.~\ref{fig:coll1} and
Fig.~\ref{fig:coll2}.  Out of the total sample, 27 systems have been or are
being followed up for further analyses (see Table~\ref{tab:folup}) and most of
them are confirmed lens systems, 5 systems are very likely lenses and rest of
the 96 are possible lenses. The Table~\ref{tab:folup} does not report
information about any archival data on any of the lens candidates. 

\begin{figure*}
\begin{center}
\includegraphics[scale=0.25]{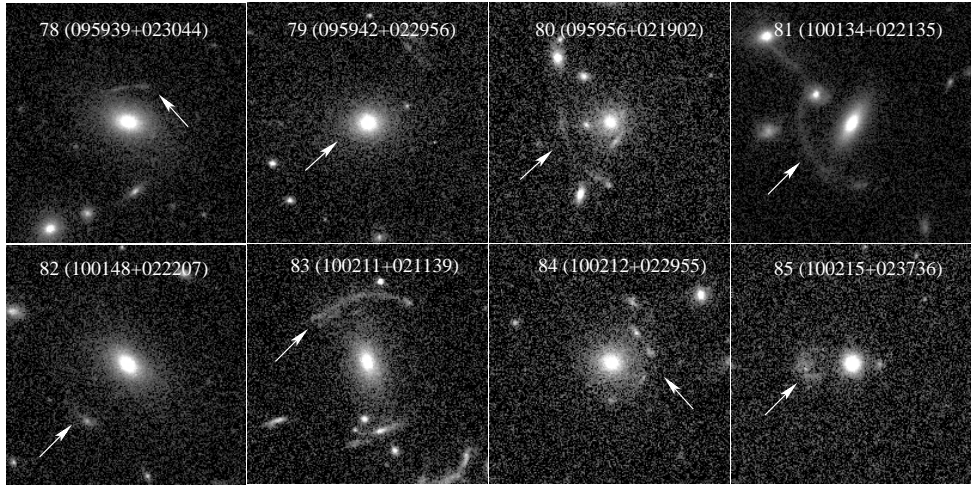}
\caption{ \label{fig:acs}
\HST\ $I$-band (F814W) imaging of the \SA lens candidates within the D2 field
extracted from the COSMOS archive. The IDs and co-ordinates are labelled for
each system. The arrows point towards the putative lensed arcs. The arc in SA79
is too faint to be visible in the $I$-band \HST\ image shown here. All
candidates are new detections except SA78 (COSMOS ID - 5939+3044) and SA83
(COSMOS ID - 0211+1139). The images are 6\arcsec on the side.  North is up and
East is left.}
\end{center}
\end{figure*}

One of the deep fields, D2, has an overlap with the COSMOS \citep{scoville07},
a photometric survey of 1.8~deg$^2$ in the $I$ band (limiting magnitude of 26.5)
with the Advanced Camera Survey (ACS) on \HST. We find two of the \SA candidates
common to the lens sample of COSMOS \citep{faure08}. The \SA candidates with IDs
SA78 and SA83 are known COSMOS candidates and the rest of the 6 candidates
within the COSMOS field are new (see Table~\ref{tab:all} with Field labelled as
D2). For comparison, we also show the ACS images of the \SA candidates in
Fig.~\ref{fig:acs} from the COSMOS archive. The W3 field overlaps with one of
the 22 individual fields of the RCS survey which covers a combined area of
$\sim$90~deg$^2$. The overlapping RCS field has a lensing cluster, RCS 1419.2+5326 
at $z=0.64$. This cluster is also detected in the \SA sample
identified by SA102 (see Table~\ref{tab:all}). 


\begin{longtable*}{l c c c c c c c c c c l}
\caption{ \label{tab:all} 
Characteristics of lens candidates from the \SA sample. }\\
\hline
\hline
ID & RA & Dec & $g$ & $r$ & $i$ & $z_{\rm phot}$ & \multicolumn{2}{c}{$R_{\rm A}$} & Rank & Type & Field \\
    & hms & dms & mag & mag & mag &   & \arcsec & $h^{-1}$Kpc &  &  & \\ [1.5ex] \hline \\ 
\endfirsthead

\hline
\hline
ID & Ra & Dec & $g$ & $r$ & $i$ & $z_{\rm phot}$ & \multicolumn{2}{c}{$R_{\rm A}$} & Rank & Type & Field \\
    & hms & dms & mag & mag & mag &   & \arcsec & $h^{-1}$Kpc &  &  & \\ [1.5ex] \hline \\ 
\endhead

\hline  
\multicolumn{12}{p{15cm}}{
{\bf m:} This galaxy falls within the masked region as per the catalog from
which the magnitudes and the redshift are extracted.  {\bf z:} The magnitudes
and/or redshift are not from the Coupon et al. catalog instead are measured by the author
using sextractor and/or {\sc zebra} \citep{feldmann06}, respectively.  {\bf C:} This lens is
identified in both D2 and COSMOS fields. Note that other lenses within D2 have
not been reported in the COSMOS lens sample \citep{faure08}.  {\bf R:} This lens is also found
in the RCS (see G03).  {\bf G:} This lens is also a part of the
SL2S-RINGS sample. }\\
\endlastfoot
   SA1         & 02:01:21.89 & $-$09:15:15.09 & 22.06 & 20.58 & 19.90 & 0.46$\pm$0.02     &  2.2 &   9.0 &  2.0 & A &  W1 \\ 
   SA2         & 02:02:10.50 & $-$11:09:11.68 & 19.79 & 18.48 & 17.80 & 0.48$\pm$0.02     &  5.0 &  20.9 &  3.7 & A &  W1 \\ 
   SA3         & 02:02:38.87 & $-$06:34:56.12 & 20.89 & 19.91 & 19.54 & 0.37$\pm$0.03     &  2.2 &   7.9 &  2.0 & A &  W1 \\ 
   SA4         & 02:03:02.84 & $-$08:21:14.25 & 21.96 & 20.57 & 19.99 & 0.14$\pm$0.07     &  2.4 &   4.1 &  2.0 & V &  W1 \\ 
   SA5         & 02:03:12.61 & $-$10:47:07.95 & 22.02 & 20.55 & 19.52 & 0.62$\pm$0.03     &  3.0 &  14.3 &  2.3 & V &  W1 \\ 
   SA6         & 02:03:20.43 & $-$07:34:50.78 & 21.69 & 20.36 & 19.45 & 0.59$\pm$0.03     &  5.0 &  23.2 &  3.3 & A &  W1 \\ 
   SA7$^m$     & 02:03:49.98 & $-$09:42:53.51 & 17.84 & 16.73 & 16.18 & 0.25$\pm$0.02     &  5.0 &  13.7 &  3.3 & A &  W1 \\ 
   SA8         & 02:04:54.51 & $-$10:24:02.48 & 19.52 & 18.18 & 17.68 & 0.33$\pm$0.02     & 10.8 &  35.9 &  2.7 & A &  W1 \\ 
   SA9         & 02:05:03.15 & $-$11:05:46.63 & 21.54 & 20.17 & 19.16 & 0.62$\pm$0.03     &  3.3 &  15.7 &  3.0 & V &  W1 \\ 
  SA10         & 02:06:48.47 & $-$06:57:01.33 & 20.95 & 19.60 & 18.90 & 0.49$\pm$0.06     &  3.2 &  13.5 &  3.0 & A &  W1 \\ 
  SA11         & 02:08:15.66 & $-$07:24:57.81 & 21.98 & 20.55 & 19.49 & 0.62$\pm$0.02     &  4.3 &  20.4 &  2.0 & A &  W1 \\ 
  SA12         & 02:08:16.87 & $-$09:36:52.69 & 22.16 & 20.85 & 19.60 & 0.74$\pm$0.03     &  3.4 &  17.4 &  3.7 & A &  W1 \\ 
  SA13         & 02:08:41.61 & $-$07:01:28.07 & 19.85 & 18.70 & 18.20 & 0.29$\pm$0.03     &  3.5 &  10.7 &  2.0 & V &  W1 \\ 
  SA14         & 02:09:29.33 & $-$06:43:11.26 & 20.47 & 19.15 & 18.57 & 0.45$\pm$0.02     &  3.2 &  12.9 &  3.7 & A &  W1 \\ 
  SA15         & 02:09:57.67 & $-$03:54:57.08 & 21.41 & 19.98 & 19.27 & 0.43$\pm$0.03     &  3.9 &  15.3 &  2.3 & V &  W1 \\ 
  SA16         & 02:10:26.57 & $-$04:46:41.59 & 21.77 & 20.79 & 20.24 & 0.55$\pm$0.03     &  1.9 &   8.5 &  2.0 & A &  W1 \\ 
  SA17         & 02:10:51.59 & $-$03:52:52.64 & 21.79 & 20.84 & 19.91 & 0.73$\pm$0.04     &  1.9 &   9.7 &  2.0 & V &  W1 \\ 
  SA18         & 02:11:08.66 & $-$10:12:13.86 & 19.81 & 18.44 & 17.89 & 0.38$\pm$0.02     &  2.0 &   7.3 &  2.3 & A &  W1 \\ 
  SA19         & 02:11:18.49 & $-$04:27:29.20 & 23.13 & 22.48 & 21.43 & 1.19$\pm$0.07     &  3.5 &  20.3 &  3.3 & A &  W1 \\ 
  SA20         & 02:12:20.52 & $-$09:38:44.10 & 23.48 & 22.18 & 20.88 & 0.77$\pm$0.03     &  2.4 &  12.4 &  2.0 & A &  W1 \\ 
  SA21$^G$     & 02:13:24.52 & $-$07:43:54.82 & 24.11 & 23.73 & 23.18 & 0.80$\pm$0.16     &  2.8 &  14.7 &  4.0 & V &  W1 \\ 
  SA22         & 02:14:08.07 & $-$05:35:32.39 & 20.98 & 19.42 & 18.69 & 0.48$\pm$0.02     &  7.1 &  29.7 &  4.0 & A &  W1 \\ 
  SA23$^G$     & 02:14:11.24 & $-$04:05:02.71 & 22.11 & 20.92 & 19.88 & 0.74$\pm$0.04     &  1.9 &   9.7 &  4.0 & A &  W1 \\ 
  SA24         & 02:15:23.03 & $-$07:36:23.56 & 23.64 & 22.08 & 20.89 & 1.05$\pm$0.02     &  3.7 &  21.0 &  2.7 & A &  W1 \\ 
  SA25         & 02:15:52.36 & $-$07:21:01.32 & 21.50 & 20.06 & 19.29 & 0.48$\pm$0.02     &  2.8 &  11.7 &  2.0 & A &  W1 \\ 
  SA26$^m$     & 02:16:04.66 & $-$09:35:06.65 & 21.68 & 20.26 & 19.09 & 0.69$\pm$0.02     & 16.4 &  81.6 &  2.7 & V &  W1 \\ 
  SA27$^{mG}$  & 02:16:24.03 & $-$09:57:39.09 & 17.29 & 16.35 & 15.90 & 0.18$\pm$0.02     &  2.8 &   5.9 &  3.3 & A &  W1 \\ 
  SA28         & 02:16:31.19 & $-$07:31:57.13 & 21.90 & 20.89 & 19.80 & 0.86$\pm$0.04     &  2.4 &  12.9 &  2.3 & A &  W1 \\ 
  SA29         & 02:16:46.84 & $-$09:18:16.74 & 21.73 & 20.53 & 19.48 & 0.72$\pm$0.03     &  2.4 &  12.1 &  2.5 & V &  W1 \\ 
  SA30         & 02:16:49.25 & $-$07:03:23.80 & 21.04 & 19.54 & 18.85 & 0.45$\pm$0.02     &  5.6 &  22.6 &  3.7 & A &  W1 \\ 
  SA31$^m$     & 02:17:23.76 & $-$10:15:50.30 & 18.68 & 17.53 & 17.03 & 0.27$\pm$0.02     &  3.2 &   9.3 &  2.3 & V &  W1 \\ 
  SA32         & 02:17:39.56 & $-$10:33:19.93 & 22.94 & 22.05 & 21.04 & 1.05$\pm$0.05     &  1.9 &  10.8 &  2.3 & A &  W1 \\ 
  SA33$^m$     & 02:18:07.29 & $-$05:15:36.16 & 22.54 & 21.21 & 20.18 & 0.42$\pm$0.03     &  2.4 &   9.3 &  3.7 & V &  W1 \\ 
  SA34         & 02:18:14.39 & $-$10:06:02.30 & 21.20 & 20.57 & 20.27 & 0.46$\pm$0.03     &  2.8 &  11.4 &  2.0 & A &  W1 \\ 
  SA35         & 02:19:09.86 & $-$04:01:43.32 & 21.43 & 19.95 & 19.27 & 0.45$\pm$0.02     &  4.3 &  17.3 &  2.3 & A &  W1 \\ 
  SA36         & 02:19:56.42 & $-$05:27:59.21 & 20.48 & 19.38 & 18.84 & 0.35$\pm$0.04     &  3.0 &  10.4 &  3.0 & V &  W1 \\ 
  SA37         & 02:20:43.11 & $-$10:52:16.45 & 22.81 & 21.68 & 20.48 & 0.79$\pm$0.03     &  2.2 &  11.5 &  2.7 & A &  W1 \\ 
  SA38         & 02:20:56.43 & $-$07:43:11.71 & 22.91 & 21.56 & 20.51 & 0.71$\pm$0.03     &  2.4 &  12.1 &  2.3 & A &  W1 \\ 
  SA39         & 02:21:51.18 & $-$06:47:32.66 & 21.34 & 20.18 & 19.16 & 0.72$\pm$0.03     &  5.2 &  26.3 &  4.0 & V &  W1 \\ 
  SA40         & 02:23:15.41 & $-$06:29:06.40 & 21.20 & 20.02 & 19.21 & 0.55$\pm$0.06     &  1.9 &   8.5 &  3.0 & A &  W1 \\ 
  SA41         & 02:23:18.33 & $-$10:58:48.46 & 21.57 & 20.29 & 19.53 & 0.52$\pm$0.04     &  6.1 &  26.6 &  2.0 & A &  W1 \\ 
  SA42         & 02:24:00.92 & $-$03:46:25.83 & 23.12 & 22.04 & 20.92 & 0.98$\pm$0.05     &  2.6 &  14.5 &  2.7 & A &  W1 \\ 
  SA43         & 02:24:05.01 & $-$04:47:07.00 & 20.00 & 18.68 & 18.10 & 0.36$\pm$0.04     &  4.3 &  15.1 &  2.0 & A &  D1 \\ 
  SA44         & 02:24:34.96 & $-$04:11:35.02 & 24.01 & 23.03 & 22.06 & 0.68$\pm$0.04     &  1.9 &   9.4 &  2.7 & V &  D1 \\ 
  SA45         & 02:24:35.26 & $-$04:01:57.86 & 22.49 & 21.21 & 20.13 & 1.13$\pm$0.07     &  3.5 &  20.1 &  2.0 & V &  D1 \\ 
  SA46         & 02:24:39.06 & $-$04:00:45.16 & 20.75 & 19.33 & 18.62 & 0.43$\pm$0.05     &  3.2 &  12.6 &  3.0 & V &  D1 \\ 
  SA47         & 02:24:59.25 & $-$04:01:03.77 & 24.05 & 22.83 & 21.70 & 0.80$\pm$0.04     &  1.9 &  10.0 &  3.0 & V &  D1 \\ 
  SA48$^{zG}$  & 02:25:11.04 & $-$04:54:33.54 & 18.72 & 17.57 & 17.03 & 0.33$\pm$0.01     &  2.8 &   9.3 &  3.7 & A &  D1 \\  
  SA49         & 02:25:38.74 & $-$04:03:20.36 & 22.09 & 20.64 & 19.52 & 0.62$\pm$0.06     &  4.3 &  20.4 &  2.0 & A &  D1 \\ 
  SA50         & 02:25:46.13 & $-$07:37:38.52 & 20.99 & 19.50 & 18.60 & 0.54$\pm$0.02     &  5.8 &  25.8 &  4.0 & V &  W1 \\ 
  SA51         & 02:26:07.15 & $-$04:27:26.26 & 19.36 & 18.44 & 17.97 & 0.17$\pm$0.05     &  3.7 &   7.5 &  2.0 & V &  D1 \\ 
  SA52         & 02:27:20.22 & $-$07:49:20.19 & 21.31 & 19.81 & 19.00 & 0.53$\pm$0.03     &  2.1 &   9.3 &  2.0 & V &  W1 \\ 
  SA53         & 02:27:59.21 & $-$09:07:29.86 & 20.71 & 19.87 & 19.42 & 0.55$\pm$0.03     &  3.9 &  17.5 &  2.0 & A &  W1 \\ 
  SA54         & 02:28:32.05 & $-$09:49:45.44 & 20.25 & 18.74 & 18.07 & 0.45$\pm$0.02     &  6.3 &  25.4 &  2.7 & V &  W1 \\ 
  SA55         & 02:29:17.36 & $-$05:54:05.54 & 19.67 & 18.30 & 17.73 & 0.38$\pm$0.03     &  2.6 &   9.5 &  3.0 & A &  W1 \\ 
  SA56         & 02:30:39.96 & $-$03:50:28.06 & 22.36 & 21.89 & 21.32 & 0.27$\pm$0.02     &  2.1 &   6.1 &  2.0 & A &  W1 \\ 
  SA57         & 02:31:06.46 & $-$05:55:04.63 & 21.68 & 20.20 & 19.46 & 0.52$\pm$0.03     &  3.7 &  16.1 &  2.0 & A &  W1 \\ 
  SA58         & 02:32:23.77 & $-$08:50:38.37 & 22.05 & 20.76 & 20.09 & 0.46$\pm$0.04     &  2.6 &  10.6 &  2.0 & A &  W1 \\ 
  SA59         & 02:33:07.05 & $-$04:38:38.21 & 21.56 & 20.66 & 19.62 & 0.79$\pm$0.03     &  1.9 &   9.9 &  2.0 & A &  W1 \\ 
  SA60         & 02:35:01.61 & $-$09:58:32.76 & 21.78 & 20.24 & 19.07 & 0.70$\pm$0.03     &  4.7 &  23.5 &  2.7 & A &  W1 \\ 
  SA61         & 08:48:23.66 & $-$04:07:15.29 & 21.17 & 19.63 & 18.85 & 0.51$\pm$0.02     &  7.4 &  32.0 &  2.7 & A &  W2 \\ 
  SA62         & 08:50:07.72 & $-$01:23:53.30 & 22.23 & 20.92 & 20.35 & 0.37$\pm$0.04     &  3.5 &  12.5 &  2.7 & A &  W2 \\ 
  SA63         & 08:52:07.18 & $-$03:43:16.28 & 20.91 & 19.33 & 18.61 & 0.48$\pm$0.02     &  5.0 &  20.9 &  3.3 & V &  W2 \\ 
  SA64         & 08:52:08.36 & $-$04:05:28.36 & 21.62 & 20.26 & 19.59 & 0.43$\pm$0.03     &  2.4 &   9.4 &  2.0 & A &  W2 \\ 
  SA65         & 08:54:25.14 & $-$03:14:53.11 & 24.89 & 24.53 & 22.84 & 0.98$\pm$0.10     &  1.9 &  10.6 &  2.7 & A &  W2 \\ 
  SA66$^m$     & 08:54:46.55 & $-$01:21:37.08 & 19.34 & 17.88 & 17.28 & 0.48$\pm$0.11     &  4.8 &  20.1 &  4.0 & A &  W2 \\ 
  SA67         & 08:55:59.92 & $-$04:09:17.76 & 21.06 & 19.60 & 18.90 & 0.45$\pm$0.02     &  2.1 &   8.5 &  3.0 & A &  W2 \\ 
  SA68         & 08:57:26.91 & $-$02:42:26.64 & 19.91 & 18.44 & 17.80 & 0.42$\pm$0.02     &  2.8 &  10.8 &  2.3 & A &  W2 \\ 
  SA69         & 08:57:35.96 & $-$01:01:12.55 & 21.03 & 20.57 & 20.29 & 0.05$\pm$0.23     &  2.4 &   1.6 &  2.3 & A &  W2 \\ 
  SA70         & 08:57:49.10 & $-$01:13:00.73 & 19.99 & 18.81 & 18.26 & 0.29$\pm$0.03     &  3.9 &  11.9 &  2.0 & A &  W2 \\ 
  SA71$^m$     & 08:58:48.83 & $-$02:39:25.79 & 19.16 & 18.23 & 17.83 & 0.36$\pm$0.10     &  3.7 &  13.0 &  3.0 & V &  W2 \\ 
  SA72         & 08:59:14.55 & $-$03:45:14.85 & 22.01 & 20.75 & 19.67 & 0.74$\pm$0.03     &  4.5 &  23.0 &  4.0 & A &  W2 \\ 
  SA73$^m$     & 08:59:54.54 & $-$01:32:13.39 & 20.87 & 19.47 & 18.85 & 0.66$\pm$1.06     &  4.3 &  21.0 &  2.0 & A &  W2 \\ 
  SA74         & 09:00:50.10 & $-$02:30:54.15 & 20.52 & 19.23 & 18.65 & 0.36$\pm$0.02     &  3.2 &  11.3 &  2.0 & V &  W2 \\ 
  SA75         & 09:02:20.42 & $-$02:30:57.28 & 22.76 & 21.97 & 21.41 & 0.63$\pm$0.04     &  3.0 &  14.4 &  2.3 & A &  W2 \\ 
  SA76$^G$     & 09:04:07.97 & $-$00:59:52.85 & 22.03 & 21.18 & 20.22 & 0.77$\pm$0.04     &  2.4 &  12.4 &  3.3 & A &  W2 \\ 
  SA77         & 09:05:29.75 & $-$02:03:17.70 & 20.94 & 19.47 & 18.80 & 0.42$\pm$0.02     &  2.4 &   9.3 &  2.0 & A &  W2 \\ 
  SA78$^C$     & 09:59:39.17 &   +02:30:43.98 & 22.72 & 21.20 & 19.92 & 0.74$\pm$0.06     &  3.2 &  16.4 &  3.0 & V &  D2 \\ 
  SA79         & 09:59:42.43 &   +02:29:56.10 & 23.10 & 21.64 & 20.32 & 0.76$\pm$0.04     &  3.5 &  18.1 &  3.0 & V &  D2 \\ 
  SA80         & 09:59:55.98 &   +02:19:01.79 & 23.30 & 22.19 & 21.06 & 1.00$\pm$0.04     &  2.4 &  13.5 &  3.3 & A &  D2 \\ 
  SA81         & 10:01:33.74 &   +02:21:35.35 & 21.96 & 20.71 & 20.08 & 0.69$\pm$0.05     &  3.0 &  14.9 &  2.0 & V &  D2 \\ 
  SA82         & 10:01:47.79 &   +02:22:06.55 & 22.92 & 21.44 & 20.20 & 0.69$\pm$0.05     &  3.5 &  17.4 &  2.7 & V &  D2 \\ 
  SA83$^C$     & 10:02:11.22 &   +02:11:39.46 & 23.64 & 22.02 & 20.77 & 0.89$\pm$0.05     &  2.6 &  14.1 &  4.0 & V &  D2 \\ 
  SA84         & 10:02:11.67 &   +02:29:55.24 & 22.81 & 21.55 & 20.56 & 0.77$\pm$0.05     &  1.9 &   9.9 &  3.0 & A &  D2 \\ 
  SA85         & 10:02:14.85 &   +02:37:36.47 & 23.51 & 22.20 & 21.15 & 0.65$\pm$0.05     &  2.1 &  10.2 &  3.0 & V &  D2 \\ 
  SA86         & 13:56:49.33 &   +55:27:07.00 & 20.41 & 18.86 & 18.21 & 0.46$\pm$0.03     &  3.7 &  15.1 &  2.5 & A &  W3 \\ 
  SA87         & 13:57:25.48 &   +53:17:43.96 & 20.52 & 19.03 & 18.15 & 0.54$\pm$0.02     &  3.5 &  15.6 &  3.0 & A &  W3 \\ 
  SA88         & 13:59:47.26 &   +55:35:37.57 & 23.40 & 22.63 & 21.57 & 0.87$\pm$0.04     &  2.2 &  11.9 &  2.3 & A &  W3 \\ 
  SA89         & 14:00:40.17 &   +56:07:49.41 & 20.47 & 19.08 & 18.48 & 0.42$\pm$0.03     &  3.7 &  14.3 &  3.0 & A &  W3 \\ 
  SA90         & 14:01:10.46 &   +56:54:20.51 & 20.89 & 19.42 & 18.57 & 0.53$\pm$0.03     &  3.7 &  16.3 &  3.7 & A &  W3 \\ 
  SA91         & 14:01:44.90 &   +53:02:09.62 & 22.01 & 20.50 & 19.61 & 0.56$\pm$0.03     &  3.0 &  13.6 &  3.7 & A &  W3 \\ 
  SA92$^G$     & 14:01:56.39 &   +55:44:46.78 & 20.79 & 19.40 & 18.70 & 0.50$\pm$0.03     &  2.8 &  12.0 &  2.0 & V &  W3 \\ 
  SA93         & 14:02:47.90 &   +57:08:52.04 & 23.70 & 22.75 & 21.68 & 1.22$\pm$0.04     &  3.2 &  18.6 &  2.0 & A &  W3 \\ 
  SA94         & 14:03:51.68 &   +57:23:50.41 & 24.08 & 22.48 & 21.72 & 0.51$\pm$0.03     &  0.0 &   0.0 &  2.7 & A &  W3 \\ 
  SA95$^G$     & 14:04:54.46 &   +52:00:24.70 & 20.10 & 18.66 & 17.88 & 0.49$\pm$0.03     &  2.2 &   9.3 &  2.7 & A &  W3 \\ 
  SA96$^m$     & 14:05:54.33 &   +54:45:48.68 & 19.50 & 18.18 & 17.49 & 0.41$\pm$0.03     &  2.8 &  10.7 &  3.0 & V &  W3 \\ 
  SA97         & 14:08:13.82 &   +54:29:08.12 & 20.28 & 18.79 & 18.04 & 0.48$\pm$0.02     &  8.0 &  33.4 &  4.0 & A &  W3 \\ 
  SA98         & 14:11:20.53 &   +52:12:09.91 & 20.16 & 18.75 & 17.93 & 0.52$\pm$0.03     & 18.4 &  80.3 &  2.0 & A &  W3 \\ 
  SA99$^m$     & 14:13:55.43 &   +53:43:44.72 & 19.03 & 17.83 & 17.26 & 0.29$\pm$0.03     &  2.4 &   7.3 &  2.0 & A &  W3 \\ 
 SA100         & 14:14:47.19 &   +54:47:03.59 & 21.19 & 19.67 & 18.45 & 0.63$\pm$0.02     & 14.7 &  70.3 &  3.7 & A &  W3 \\ 
 SA101         & 14:16:44.52 &   +56:42:16.18 & 22.99 & 21.41 & 19.94 & 1.29$\pm$0.16     &  3.5 &  20.5 &  2.0 & A &  W3 \\ 
 SA102$^R$     & 14:19:12.17 &   +53:26:11.44 & 21.83 & 20.30 & 19.11 & 0.69$\pm$0.02     &  9.9 &  49.2 &  3.7 & A &  W3 \\ 
 SA103$^{mG}$  & 14:19:17.25 &   +51:17:28.63 & 20.78 & 19.50 & 18.72 & 0.47$\pm$0.03     &  4.1 &  16.9 &  3.0 & V &  W3 \\ 
 SA104         & 14:21:02.56 &   +52:29:42.51 & 17.74 & 16.79 & 16.33 & 0.18$\pm$0.01$^z$ & 11.7 &  24.9 &  2.0 & A &  D3 \\
 SA105         & 14:21:18.35 &   +52:50:22.37 & 21.40 & 19.89 & 19.14 & 0.47$\pm$0.05     &  2.8 &  11.6 &  3.0 & V &  D3 \\ 
 SA106         & 14:22:58.34 &   +51:24:39.50 & 22.78 & 21.75 & 20.80 & 0.74$\pm$0.04     &  1.9 &   9.7 &  2.5 & A &  W3 \\ 
 SA107         & 14:23:49.27 &   +57:26:33.90 & 23.20 & 22.15 & 21.21 & 0.69$\pm$0.06     &  2.2 &  10.9 &  2.0 & A &  W3 \\ 
 SA108         & 14:25:44.27 &   +57:07:24.47 & 25.45 & 23.87 & 22.53 & 0.86$\pm$0.04     &  4.5 &  24.2 &  2.7 & A &  W3 \\ 
 SA109         & 14:26:08.04 &   +57:45:23.90 & 20.56 & 19.49 & 18.99 & 0.39$\pm$0.03     &  3.2 &  11.9 &  2.0 & A &  W3 \\ 
 SA110$^m$     & 14:28:10.54 &   +56:39:48.36 & 17.67 & 16.76 & 16.30 & 0.80$\pm$0.28     &  4.1 &  21.5 &  2.3 & A &  W3 \\ 
 SA111         & 14:28:34.82 &   +52:13:06.44 & 22.28 & 20.75 & 19.94 & 0.52$\pm$0.03     &  5.0 &  21.8 &  2.7 & A &  W3 \\ 
 SA112         & 14:30:00.65 &   +55:46:47.97 & 21.58 & 20.02 & 19.12 & 0.55$\pm$0.02     &  4.3 &  19.3 &  4.0 & A &  W3 \\ 
 SA113         & 14:31:39.77 &   +55:33:22.81 & 21.83 & 20.57 & 19.42 & 0.71$\pm$0.03     &  3.0 &  15.1 &  3.0 & V &  W3 \\ 
 SA114         & 14:31:52.67 &   +57:28:36.73 & 22.87 & 21.52 & 20.19 & 0.83$\pm$0.03     &  3.5 &  18.6 &  2.3 & A &  W3 \\ 
 SA115         & 14:34:03.87 &   +51:21:36.07 & 19.98 & 18.66 & 18.06 & 0.39$\pm$0.02     &  2.6 &   9.6 &  2.0 & A &  W3 \\ 
 SA116         & 14:34:34.69 &   +56:59:20.17 & 21.19 & 19.63 & 18.74 & 0.57$\pm$0.02     &  4.1 &  18.7 &  2.7 & A &  W3 \\ 
 SA117$^m$     & 22:01:51.79 &   +04:10:08.42 & 18.53 & 17.73 & 17.19 & 0.43$\pm$0.04     &  7.3 &  28.7 &  2.7 & A &  W4 \\ 
 SA118$^m$     & 22:02:01.66 &   +01:47:09.57 & 18.82 & 17.63 & 17.07 & 0.30$\pm$0.02     &  5.0 &  15.6 &  3.0 & A &  W4 \\ 
 SA119$^G$     & 22:03:29.03 &   +02:05:18.89 & 21.24 & 19.99 & 19.37 & 0.38$\pm$0.04     &  2.6 &   9.5 &  4.0 & V &  W4 \\ 
 SA120$^G$     & 22:05:06.92 &   +01:47:03.71 & 21.20 & 19.91 & 19.15 & 0.46$\pm$0.06     &  2.1 &   8.6 &  2.0 & A &  W4 \\ 
 SA121         & 22:06:42.03 &   +04:11:30.85 & 21.20 & 19.81 & 18.88 & 0.62$\pm$0.03     &  3.7 &  17.6 &  3.0 & A &  W4 \\ 
 SA122         & 22:13:06.93 & $-$00:30:37.05 & 21.19 & 19.98 & 18.81 & 0.69$\pm$0.02     &  2.8 &  13.9 &  3.0 & A &  W4 \\ 
 SA123         & 22:13:31.85 &   +00:48:36.14 & 23.37 & 21.87 & 20.56 & 1.00$\pm$0.03     &  4.8 &  26.9 &  4.0 & A &  W4 \\ 
 SA124         & 22:14:09.57 & $-$17:30:56.23 & 22.63 & 21.08 & 19.85 & 0.83$\pm$0.05     &  7.4 &  39.4 &  4.0 & V &  D4 \\ 
 SA125         & 22:14:18.82 &   +01:10:33.85 & 20.31 & 19.24 & 18.84 & 0.74$\pm$0.08     &  0.0 &   0.0 &  3.0 & A &  W4 \\ 
 SA126         & 22:17:29.38 & $-$00:38:36.60 & 19.93 & 18.94 & 18.32 & 0.78$\pm$0.02     &  1.9 &   9.9 &  2.0 & A &  W4 \\ 
 SA127         & 22:21:43.74 & $-$00:53:02.89 & 19.36 & 17.98 & 17.35 & 0.39$\pm$0.02     &  4.7 &  17.4 &  3.3 & V &  W4 \\ 
\end{longtable*}


\begin{table*}
\begin{center}
\caption{ \label{tab:folup} 
{\SA candidates with follow-up information from the SL2S collaboration.} }
\begin{tabular}{ccccl}
\hline
\hline
 RA & DEC & Reference & $z_l$, $z_s$ & Comment \\
\tableline 

 02:09:57.67 &  $-$03:54:57.08  &  VM         &   --, --       &  V, H  \\ 
 02:13:24.52 &  $-$07:43:54.82  &  --         &  0.72, --      &  K, V, H \\ 
 02:14:08.07 &  $-$05:35:32.39  &  L09,V11    &  0.444,        &  V, H \\ 
       &                  &  VM         &    1.023$\pm$0.001/1.7$\pm$0.1   &  \\
 02:14:11.24 &  $-$04:05:02.71  &  M11,VM     &  0.609, --     &  K, V, H   \\ 
 02:16:49.25 &  $-$07:03:23.80  &  VM         &   --, --       &  V, H \\ 
 02:18:07.29 &  $-$05:15:36.16  &  M11        &  0.647, --     &  V, H \\ 
 02:19:56.42 &  $-$05:27:59.21  &  VM         &   --, --       &  V, H \\ 
 02:21:51.18 &  $-$06:47:32.66  &  L09,VM     &  0.618,        &  V, H\\ 
 02:25:11.04 &  $-$04:54:33.54  &  R11        &  0.238, 1.199  &  K, H \\  
 02:25:46.13 &  $-$07:37:38.52  &  L09        &  0.511,        &  G  \\ 
 08:52:07.18 &  $-$03:43:16.28  &  VM         &   --, --       &  V, H  \\ 
 08:54:46.55 &  $-$01:21:37.08  &  L09,L10    & 0.3530$\pm$0.0005,1.2680$\pm$0.0003   &  K, V, H \\
 08:58:48.83 &  $-$02:39:25.79  &  --         &   --, --       &  H \\ 
 08:59:14.55 &  $-$03:45:14.85  &  L09,VM     &  0.647, --     &  V, H\\ 
 09:04:07.97 &  $-$00:59:52.85  &  --         &   --, --       &  H \\ 
 10:02:11.22 &  +02:11:39.46    &  --         &   --, --       &  H \\ 
 14:08:13.82 &  +54:29:08.12    &  L09        &  0.416, --     &  S, H \\ 
 14:14:47.19 &  +54:47:03.59    &  --         &   --, --       &  H \\ 
 14:19:12.17 &  +53:26:11.44    &  --         &   --, --       &  H \\  
 14:19:17.25 &  +51:17:28.63    &  --         &   --, --       &  H \\  
 14:30:00.65 &  +55:46:47.97    &  T10        &  0.497$\pm$0.001, 1.435$\pm$0.001  &  G, H  \\ 
 14:31:39.77 &  +55:33:22.81    &  T10        &  0.669$\pm$0.001, --     &  G, H  \\ 
 22:03:29.03 &  +02:05:18.89    &  --         &   --, --       &  H \\   
 22:13:06.93 &  $-$00:30:37.05  &  L09        &   --, --       &  H \\    
 22:13:31.85 &  +00:48:36.14    &  L09        &   --, --       &  H \\  
 22:14:18.82 &  +01:10:33.85    &  --         &   --, --       &  H \\    
 22:21:43.74 &  $-$00:53:02.89  &  L09        &  0.334, --     &  S, H\\ 
\tableline
\end{tabular}
\tablecomments{
{\bf Col 3:} L09-\cite{limousin09}, L10-\cite{limousin10},
T10-\cite{thanjavur10}, R11- \cite{ruff11}, V11-\cite{verdugo11} and M11-Mu\~noz
et al. (2011, in prep, PI- V. Motta, ESO-080.A-0610). VM is assigned to those systems which are being followed
up for spectroscopy with the VLT by V. Motta (PI, ESO-086.A-0412). {\bf  Col 4:}
Spectroscopic redshifts for the lens galaxy and lensed source from the follow-up
observations along with their error bars, if available. {\bf Col 5:} V-VLT, K-Keck or G-Gemini indicates telescopes used
for follow-up spectroscopy and S-SDSS spectroscopy. H corresponds to HST imaging
followed up by J-P. Kneib (PI, C15 and C16) and/or R. Gavazzi (PI, C17). }
\end{center}
\end{table*}


\section{RESULTS AND DISCUSSION}
\label{sec:resdisc}

In the following subsections, we describe the main findings from the \SA sample
and constraints on average properties of the lens population using statistical
properties of the arcs. Note that many of the lens candidates are not confirmed
lenses yet and hence, the results should be taken as indicative. Firstly, based
on the photometric redshifts of the lenses, we study the lens redshift
distribution. Subsequently, we discuss about the abundance of giant arcs and
presence of radial arcs in the sample. Finally, we measure the azimuthal
distribution and image separation distribution of the arcs and argue their
importance as diagnostics for understanding the matter distribution of the
lenses, statistically.


\subsection{Lens redshift distribution}

\begin{figure}
\begin{center}
\includegraphics[scale=0.85]{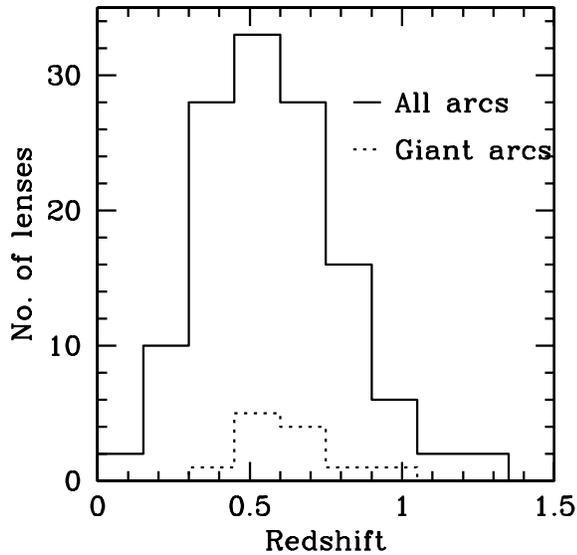}
\caption{ \label{fig:lenshist}
Redshift distributions of the lens galaxies from the \SA sample. The solid
histogram corresponds to lenses from the whole sample whereas the dashed histogram
corresponds to lenses with giant arcs only. The respective means of their redshift
distributions are $z=0.58\pm0.22$ and $z=0.64\pm0.19$ where the uncertainties 
indicate 68\% confidence level.}
\end{center}
\end{figure}

The lenses producing giant arcs (e.g. length-to-width ratio $\ge10$) consist of clusters
which lie at the high end of the halo mass function. The redshift distribution
and abundance of such massive clusters depend on the parameters of a
cosmological model. \citet{bartelmann98} estimated that most of the lensing
clusters giving rise to giant arcs should peak at around $z\sim0.3-0.4$ for the
currently accepted standard cosmological model.

We use the \CF photometric redshift catalogs \citep{coupon09} which are
generated from the code {\sc le phare} \citep{ilbert06} to calculate the \SA
lens redshift distribution. The accuracy on the redshifts of galaxies in the \WD
with magnitudes $i<21.5$ is $\sigma_{\Delta z/(1+z)}~\sim~0.037$ and with
magnitudes $22.5 <i<23.5$ is $\sigma_{\Delta z/(1+z)}~\sim~0.08$. In
Fig.~\ref{fig:lenshist}, we show the redshift distribution for all the lenses in
the \SA sample (solid histogram) and for lenses consisting of giant arcs only
(dashed histogram). In the appendix B, we describe how we estimated the mean and
1$\sigma$ uncertainty given the redshift measurement errors (see
Table~\ref{tab:all}). We find that the mean of the lens redshift distribution
for the \SA sample is $z=0.58\pm0.22$ and that for the sample of giant arcs is
$z=0.64\pm0.19$. We note that the mean of the giant arcs sample is at a
higher redshift compared to the peak expected from \citet{bartelmann98} but
certainly consistent within 2$\sigma$ confidence interval. For comparison,
the RCS sample G03 finds that most of the lenses with giant arcs have
redshifts upwards of $\sim0.6$. 

We make a qualitative comparison of the redshift distributions of the lens
populations from other surveys in the literature (not shown in the figure).
However, we note that these surveys have significantly different selection
functions and hence, quantitative conclusions should not be drawn. The
distribution of lens sample from
CASTLES\footnote{http://www.cfa.harvard.edu/castles/} peaks between 0.3 and 0.4
which is lower than the mean redshift of the \SA sample. CASTLES has a large
enough sample of lenses but is not homogeneously selected. Nevertheless, the
peaks are consistent within 2$\sigma$ (assuming an error of 0.05) in spite of
the differences in the sample selection. The COSMOS lens sample
\citep{faure08}, on the other hand, is fairly homogeneous but the sample size
of confirmed lenses is relatively small. COSMOS has a bimodal lens redshift
distribution with a minimum at $\sim 0.5-0.6$. This is in stark contrast with
the redshift distribution of \SA (see Fig.~\ref{fig:lenshist}). 


\begin{figure*}
\begin{center}
\includegraphics[scale=0.55]{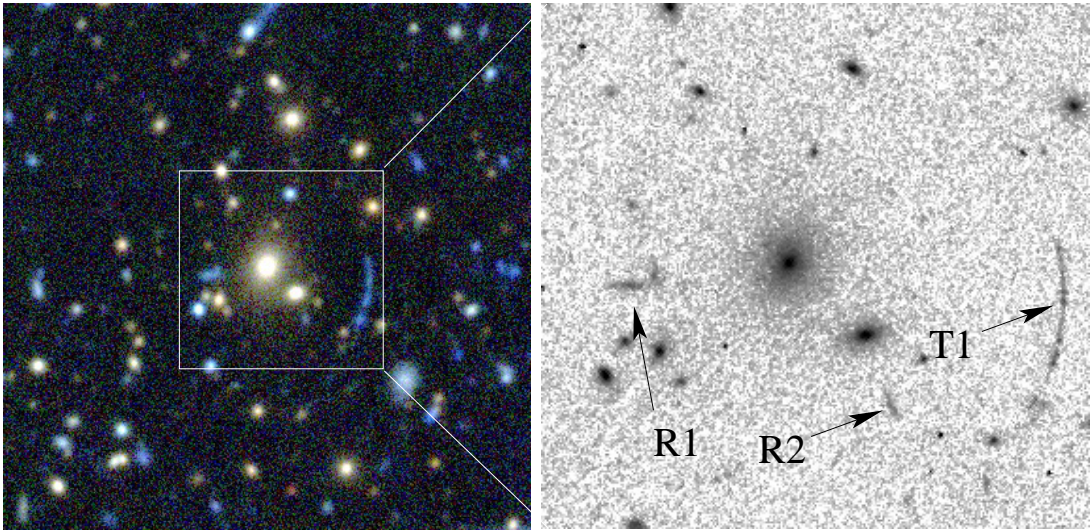}
\caption{ \label{fig:radarc}
The \CF $gri$ image of SL2SJ141447+544704 (SA100) on the left and \HST
\ F606W ($V$ band) image on the right. The high resolution of \HST \ suggests two radial arc
candidates labeled as R1 and R2 whereas the tangential arc labeled as T1 is a
confirmed lensed arc. The \CF image is 74.4\arcsec on the side and the \HST \ image is
$\sim$28.3\arcsec on the side.}
\end{center}
\end{figure*}

\subsection{Giant and radial arcs}
\label{sec:grarcs}
Here, we report detections of giant and radial arcs from our sample. The arcs,
both radial and tangential, produced in massive clusters are conventionally
referred to as giant arcs, if their length-to-width ($l/w$) ratio is larger than
about 8 or 10. For a (non-singular) circularly symmetric projected density profile, the
inverse magnification matrix of a lensed image has two eigenvectors, one in the
radial and the other in the tangential direction. If the radial eigenvalue
becomes 0, then the arcs are magnified and stretched radially with respect to
the center of the lens and are called radial arcs. If the tangential eigenvalue
becomes 0, then the arcs are magnified and stretched tangentially and are called
tangential arcs. 

\subsubsection{Giant arcs}
The statistics of giant arcs allow detection of clusters at the massive end of
halo mass function that are rare. The statistics of such rare massive
structures is sensitive to the cosmological model of the Universe.  Hence,
several attempts have been made to predict the giant arcs statistics
\citep[e.g.,][henceforth, D04]{bartelmann98,wambsganss04,dalal04} that
suggested a large discrepancy compared to the observed abundance of arcs from a
well-defined cluster population
\citep[e.g.,][]{luppino99,gonzalez01,gladders03,li05,li06}. However, the discrepancy has
been substantially diminished due to more realistic assumptions such as using a
realistic source redshift distribution and improved predictions which consider
the contribution of central galaxy or substructure within the halos
\citep[e.g.,][]{meneghetti00,meneghetti03,horesh05}.

We present the abundance of giant arcs in our sample which could be tested with
predictions from realistic simulations by taking into account the observational
limitations. Within a total unmasked area of $\sim$150~deg$^2$, 8 of the arcs
have $l/w$ of $\sim$10 or above. Additionally, 4 more arcs have $l/w~\sim~8$
which are included in the sample of giant arcs because these appear to be
broken either due to noise or due to a satellite galaxy. Since one of the giant
arcs is from the DEEP, we use rest of the 11 giant arcs from the \WD data for
comparison with RCS. We use the primary sample of RCS which has roughly similar
depth compared to \WD data. The primary RCS lens sample has 6 arcs with
$l/w\ge8$ found within a total area of 90~deg$^2$ G03. Therefore, the RCS
sample has $0.07\pm0.03$~arcs~deg$^{-2}$ which is consistent with the
$0.07\pm0.02$~arcs~deg$^{-2}$ from the \SA sample.

\subsubsection{Radial arcs}
The radial arcs are formed when the source falls on the radial caustic (see
Appendix A). The size of the radial caustic and hence, the cross-section to form
radial arcs depends on the slope of inner regions of the density profile of the
lens \citep[e.g.,][]{hattori99}. The statistics of radial arcs can thus, be
used to constrain the slope of the central density profiles of clusters and
thereby, allow a better understanding of the nature of dark matter
\citep{molikawa01}.  More thorough investigations have been carried out to test
the effects of realistic assumptions of properties such as lens ellipticity,
source size and ellipticity on the statistics of radial and tangential arcs
\citep[e.g.,][]{keeton01,oguri02b}.  \cite{sand05} used archival \HST \ (WFPC2) data on
various lensing cluster samples to do a systematic study of the number ratio of
radial to tangential arc statistics. They accounted for the effects due to the
central galaxy in the expected number ratio and placed loose constraints on the
slope of the average inner density profile of the dark matter. They underlined
the need of larger observational datasets to further explore effects of
substructure, mass of the central galaxy and homogeneity of the sample.

We report detections of radial arc candidates in the \SA sample during the
visual inspection. We found 1 candidate which appears like a radial arc in the
system SL2SJ141447+544704 (SA100 in Table~\ref{tab:all}). This lens system also
has a giant tangential arc (T1) with the same color as the radial arc (R1, see
the left panel of Fig.~\ref{fig:radarc}). The subsequent \HST \ observation in
the F606W band shows another radial arc candidate (R2, see right panel of
Fig.~\ref{fig:radarc}). The nature of the radial arc candidates needs to be
verified with spectroscopy and mass modeling since these could be blue edge-on
disk galaxies. Radial arcs are usually faint and are overshadowed by the bright
central galaxies near which they are formed. Nevertheless, we argue that a
follow-up imaging of the promising \SA candidates at high resolution will help
in creating a homogeneous sample of radial and tangential arcs which could
provide crucial insights in understanding the density distributions in the inner
regions of the lens systems.


\subsection{Galaxy-arcs orientation}

In this section, we quantify the angular distribution of arcs with respect to
the major axis of the ellipticity of the lensing halo. D04 suggested that
triaxial dark matter halos, when acting as lenses, usually form lip caustics
\citep[e.g.,][]{hattori99} and the formation of giant arcs tends to be at the
ends of the lip caustics. Using numerical simulations, they showed that the
giant arcs are oriented very close to the major axis of the dark matter halo. If
a central galaxy is further added to the halo, then it appeared to isotropize
the angular distribution of arcs to a small extent. In order to compare these
predictions to the observations, D04 measured the angular distribution of giant
arcs from the EMSS cluster sample under the assumption that the ellipticity of
the lensing galaxy is a good representation of the ellipticity of the underlying
DM halo.  They found that most of the giant arcs had an orientation of
$<$~45~deg consistent with their predictions. 

We use the same definition as that used in D04 for the orientation of the arcs,
that is, the angle between the major axis of the lens galaxy and the line
connecting the center of the lens galaxy to the midpoint of the arc. The
measurement of orientation of real arcs tends to be somewhat subjective since
there exists an ambiguity in defining the extent of the arc which is required to
find the midpoint of the arc. The midpoint of the arc and its orientation from
the major axis of the lens is calculated manually. When there is a single
dominant lensing galaxy we measure the position angle (PA) of the major axis of
the lens galaxy with {\sc sextractor} \citep{bertin96} otherwise we adopt the
following strategy. If there are two or three (nearly collinear) lens galaxies
with similar colors and brightness, then the PA of the ellipticity is assumed to
be along the line joining the lens galaxies. If a circle going through the arc
encloses multiple lens galaxies with comparable brightness but no obvious
elongation, then such system is rejected. If one of the lens galaxies is
predominantly brighter than the companion galaxies, then the ellipticity of the
dominant galaxy is assumed to reflect the overall ellipticity of the
gravitational potential and hence, of the caustics. 

\begin{figure*}
\begin{center}
\includegraphics[scale=0.75]{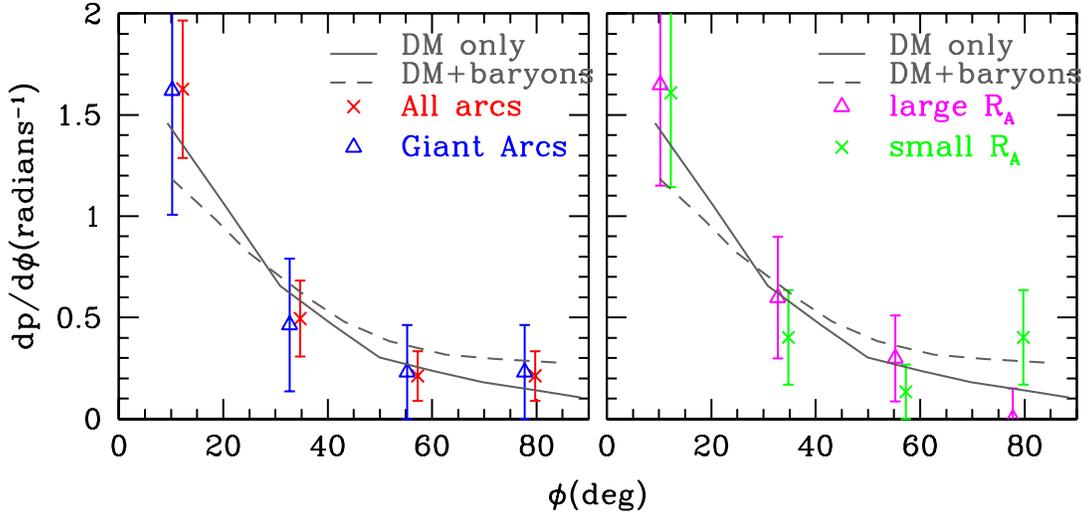}
\caption{ \label{fig:orien}
Angular distribution of arcs with respect to the lens galaxy. The
data points in both the panels are from the \SA sample. In the right panel, the ``all arcs" sample
is divided into small and large $R_{\rm A}$. The expected curves for dark matter
(DM) only (solid) and DM+$3\times10^{12}h^{-1}M_{\odot}$ (dashed) are taken from D04 which
are calculated from the cluster sample of GIF simulation \citep{kauffmann99}. }
\end{center}
\end{figure*}

After applying the above selection cuts, we are left with 36 all arc candidates
and 11 giant arcs with an average ranking $>2.5$. The angular distribution of
the giant arcs from the \SA sample is shown in the left panel of
Fig.~\ref{fig:orien}. We compare the observed distribution to the expected
distributions from D04 which correspond to two cases: a) for dark matter only
lensing halo (solid curve) and b) a dark matter halo to which a galaxy of
$3\times10^{12}h^{-1}M_{\odot}$ is added (dashed curve). The distribution of
giant arcs from our sample appears to follow the expected trend although it is
not possible to distinguish between the two cases. We also show the
distribution of orientation of all arc candidates (see left panel of
Fig.~\ref{fig:orien}) which seems to follow the expected anisotropy. This
extends the result obtained by D04 to groups-scale lens systems. Furthermore,
we split the all arcs sample into a small $R_{\rm A}$ ($<5\arcsec$) and large
$R_{\rm A}$ ($>=5\arcsec$) samples to see the dependence of the angular
distribution on the Einstein radius (or the arc radius). The choice of cut at
5\arcsec is arbitrary. The baryonic matter tends to make the matter
distribution in halos more spherical at the center. Therefore, the angular
distribution of arcs for the sample with smaller $R_{\rm A}$ is expected to be
more isotropic compared to the sample with large $R_{\rm A}$. However, we do
not find any clear differences in the angular distribution of small and large
$R_{\rm A}$ for the \SA sample within the uncertainties (see the right panel of
Fig.~\ref{fig:orien}). We note that these measurements may have some systematic
errors due to the orientation dependence of the selection function. Ambiguities
also exist in the definition of the orientation in few cases, especially when
multiple candidate lens galaxies are involved. In addition, the current
analysis consists of lens candidates as opposed to confirmed lens systems.
Therefore, our conclusions should be treated more of a qualitative nature.


\subsection{Image separation distribution}
\label{sec:isd}

The image separation distribution (ISD) is sensitive to the halo mass, abundance
of the lens population, the mass distribution in the lens and the source
redshift. Therefore, the ISD measured from galaxy to cluster scales contains
information about the cosmological parameters and various scaling relations
between galaxy properties and halo mass. Hitherto, the ISD has been measured
either at small image separations ($\theta$) primarily, with lens samples such as CASTLES
\citep[e.g.,][]{keeton00,khare01} and CLASS \citep[e.g.,][]{kochanek01,oguri06} or at large
$\theta$ with cluster-scale lenses, for example, the MACS sample
\citep{zitrin11}. With the \SA sample, we can probe the intermediate mass regime
corresponding to group scale lenses. The \SA sample is selected based on the
presence of elongated arc-like features and is not directly biased towards
selecting a specific lens population. The ISD measured from the lens samples is
referred to as the observed ISD and the ISD calculated from various models is
referred to as the expected/predicted ISD throughout this paper.

\subsubsection{Model for the expected distribution}

In the past, the expected ISD was computed with models consisting of either
galaxy dominated \citep[e.g.,][]{turner84} or dark matter dominated lensing halos
\citep[e.g.,][]{narayan88}. However, since the density profile of dark matter is known to
be affected due to the presence of baryons at the center, the need for a more
complex model to explain the observed ISD was pointed out by \citet{keeton98}
and demonstrated by subsequent studies \cite[e.g.,][henceforth,
O06]{porciani00,kochanek01,oguri06}. We follow the framework developed in O06
and \citet{oguri02a} to calculate the expected ISD in order to compare it with
the observed ISD from the \SA sample. We also adopt the cosmology used in
O06 which constitutes of the following cosmological parameters: $\Omega_m=0.3$,
$\Omega_{\Lambda}=0.7$ and $\sigma_8=0.9$. We describe in detail how each of the
model parameters are calculated before proceeding to the comparison of various
models.

The probability for a source at redshift, $z_s$ to get lensed with image
separation greater than $\theta$ is given by
\begin{eqnarray}
P (>\theta;z_s) &=& \int_0^{z_s} \dx z_l \frac{\dx \chi}{\dx z_l} \int_{0}^{\infty} \dx M \, n(M,z_l) \nonumber \\
&\times& \sigma_{{\rm lens}}\, \Theta(M-\tilde{M})  \, ,
\label{eq:cumul}
\end{eqnarray}
where $\chi$ is the comoving distance and $z_l$ is the redshift of the lens, $M$
corresponds to the halo mass, $ n(M,z_l)$ is the halo mass function, $\tilde{M}$
is the minimum halo mass that causes an image separation equal to
$\theta$, and $\Theta$ is the Heaviside step function. The biased lensing cross
section, $\sigma_{\rm lens}$ is measured in comoving units and is given by
\begin{equation}
\label{eq:siglens}
\sigma_{\rm lens}= \pi y_r^2 \xi_0^2 B(z_s) \, ,
\end{equation}
where $y_r\xi_0$ corresponds to the radius of the outermost
caustic in the lens plane\footnote{For lens models with spherical
symmetry that are considered here, the radial caustic is the
outermost caustic and a source lying within the outermost
caustic gets strongly lensed, i.e., multiply imaged.} and it depends
on the matter density distribution around the lens. The quantity
$B(z_s)$ denotes the magnification bias which causes sources, fainter
than the limiting magnitude of the survey, to be detected in the
sample. It is the ratio of the number of sources that can be
potentially lensed into an image with luminosity $L$ to the number of
sources that have an intrinsic luminosity $L$. In general, the
magnification bias depends upon the luminosity of the source and under
the assumption of spherical symmetry, it is given by
\begin{equation}
B(z_s,L)=\frac{1}{\pi y_r^2 \, \Phi(z_s,L) \, \dx L} \int_0^{y_r} dy \, 2
\pi y \, \Phi\Big(z_s,\frac{L}{\mu(y)}\Big) \frac{\dx L}{\mu(y)} \,,
\label{eq:magbias}
\end{equation}
where $\Phi(z_s,L)$ is the true source luminosity function and
$\mu(y)$ is the lensing magnification at an angular position $y$
inside the caustic. Under the assumption of a power-law luminosity
function which does not evolve with redshift, the luminosity
dependence of the magnification bias drops out.

Differentiating both sides of Eq.~\ref{eq:cumul}, we obtain the
differential probability
\begin{eqnarray}
\label{eq:isdog}
\bigg|\frac{ \dx P}{ \dx \theta}\bigg| = \int_0^{z_s} \dx z_l \frac{{\rm
d}\chi}{\dx z_l} \int_{0}^{\infty} \frac{\dx M}{\dx \theta} \, n(M,z_l)
\, \sigma_{{\rm lens}} \, \delta[M-\tilde{M}(\theta)] \nonumber \\
                  = \int_0^{z_s} \dx z_l \, \frac{{\rm
d}\chi}{\dx z_l}\, n(\tilde{M}[\theta],z_l)
\, \sigma_{{\rm lens}} \frac{\dx \tilde{M}}{\dx \theta} \,,
\end{eqnarray}
which can be directly related to the ISD of the observed lens sample.

In O06, the above equation is used to predict the ISD resulting from different
components in a given halo. To use the above equation, we have to specify the
distribution of mass inside a halo and the halo mass function. The former allows
us to calculate $\tilde{M}(\theta)$ and the biased cross section for lensing.  For the
latter, O06 assumed the then state-of-the-art calibration of the mass function
given by \citet{sheth99}. For the mass distribution inside a halo, O06
considered the following different components.

\begin{itemize}
\item At the center of the dark matter halo, the matter density is dominated by the
central galaxy and the total matter distribution is very close to that of a
singular isothermal sphere (SIS). This distribution is given in terms of the
one-dimensional velocity dispersion ($\sigma_{\rm vel}$) such that
$\rho(r)~\propto~\sigma^2_{\rm vel}/r^2$.  Following Sect 2.3.1 of
\citet{oguri02a}, we calculate the size of the radial caustic ($y_r\xi_0$) in comoving
coordinates, the total magnification of the lensed images and the relation
between the velocity dispersion and the image separation for the SIS mass
distribution. To relate the mass of the halo to the velocity dispersion of the
central galaxy, O06 used galaxy scaling relations. They adopted the halo
mass-luminosity relation from \citet{vale04} using the abundance matching
technique and the Faber-Jackson relation obtained by \citet{bernardi03} using
SDSS galaxies to relate the luminosity to the velocity dispersion of the galaxy.
Following \citet{cooray05}, they also included a log-normal scatter in the halo
mass-luminosity relation with a scatter of 0.25~dex \citep[see
also][]{more09b,more11b}.

\item On large scales, the distribution of dark matter follows the Navarro-Frenk-White
\citep[NFW,][]{navarro97} profile given by 
\begin{equation}
\rho(r) = \frac{\rho_s}{(r/r_s)(1+r/r_s)^2} \, .
\end{equation}
where the scale radius $r_s=R_{\rm vir}/c$, the concentration
parameter $c$ is related to the mass $M$ with a considerable scatter.
O06 used the mean relation between concentration and mass of \citet{bullock01}, given by
\begin{eqnarray}
\bar{c}= \frac{10}{1+z} \left(\frac{M}{M_{*}}\right)^{-0.13} \, .
\end{eqnarray}
The distribution of concentrations at fixed halo mass is assumed to be
log-normal with a scatter of $0.3$ dex.  For the NFW case, we calculate the
caustic size, the magnification and the relation between halo mass and the image
separation numerically. We defer the details of our procedure to calculate these
quantities to the appendix A.

\begin{figure}
\begin{center}
\includegraphics[scale=0.85]{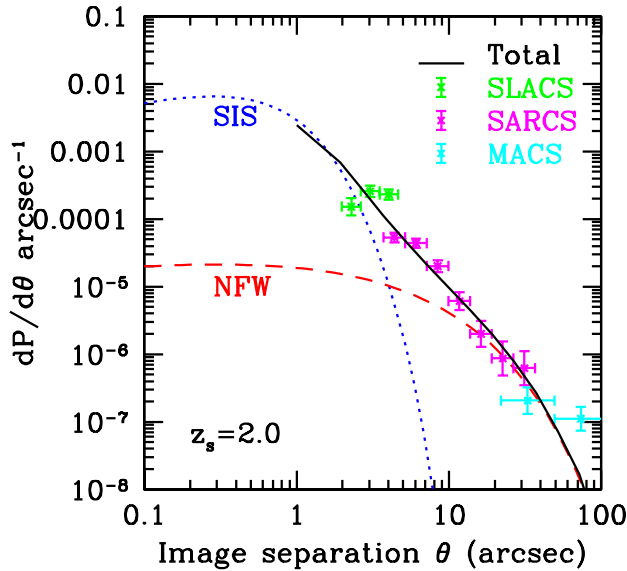}
\caption{ \label{fig:isdog} Image separation distribution. Theoretically
calculated image separation distribution curves for SIS profile (dotted), NFW
profile (dashed) and total profile (solid) following the O06 model. The data
points from the SLACS (green), \SA sample (magenta) and MACS (cyan) where the vertical bars
indicate Poissonian errors and horizontal bars show the bin width. As discussed
in Sect.~\ref{sub:obsisd}, we multiply the ISDs of the lens samples by their
respective $P(>\theta_{\rm cut})$.}
\end{center} 
\end{figure}

\item In addition to the above two simple profiles, O06 also considered a
combined total profile which included the central galaxy, the dark
matter halo and the effect of adiabatic contraction (AC) of
dark matter in response to the baryonic component of the galaxy at the
center. In this case, O06 assumed the central galaxy to have a
Hernquist profile \citep{hernquist90}, given by
\begin{eqnarray}
\rho(r) = \frac{M_b}{2\pi}\frac{1}{(r/r_b)(r_b+r)^3} \, .
\end{eqnarray}
where $M_b$ is the stellar mass of the galaxy and $r_b$ is a core
radius. The stellar mass $M_b$ was obtained using the halo mass-luminosity 
relation found by \citet{vale04} and adopting a constant
mass-to-light ratio of $3.0h_{70}M_\odot/L_\odot$. The scaling relations
of \citet{bernardi03} can be used to obtain the effective radius $R_0$
as a function of the luminosity of the galaxy which is related to the
core radius of the Hernquist profile such that $r_b=0.551R_0$. The
AC is carried out using the analytical formalism
presented by \citet{gnedin04}. Having specified the total dark matter
distribution, the caustic size, the magnification and the relation
between halo mass and the image separation need to be calculated
numerically. We describe the procedure we use, in the appendix A.
\end{itemize}

\begin{figure*}
\begin{center}
\includegraphics[scale=0.85]{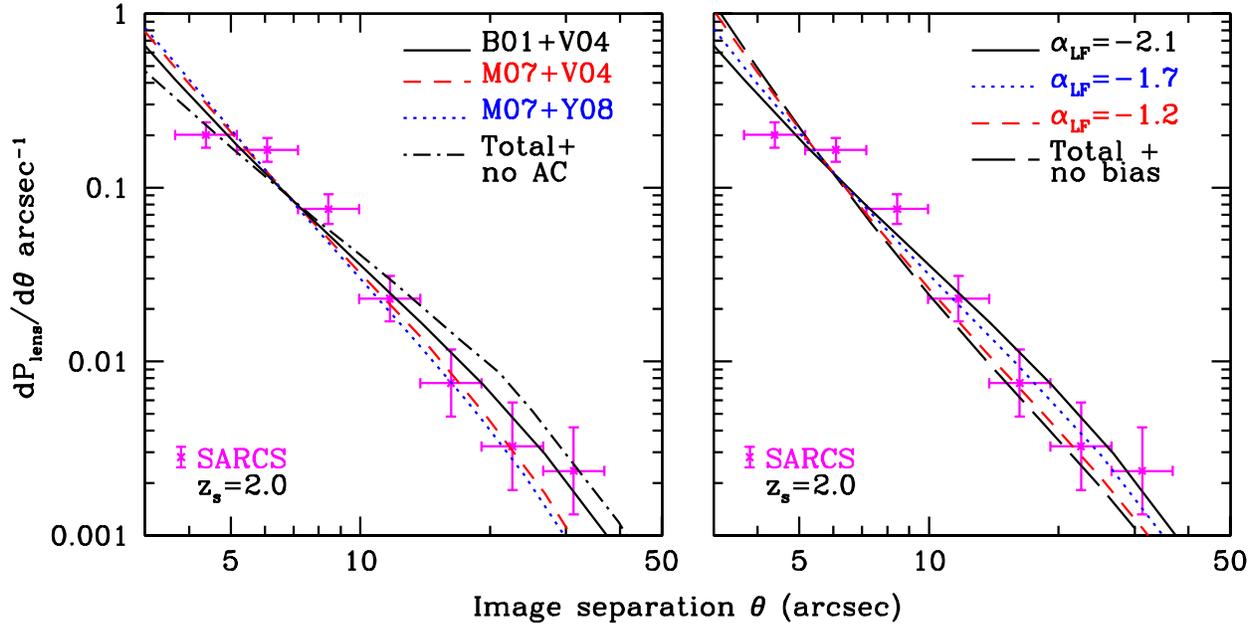}
\caption{ \label{fig:isdts} 
Image separation distribution. Theoretical curves for the total
profile, shown with solid black line, are the same as in Fig.~\ref{fig:isdog}.
{\it Left:} Adopting the $c$-$M$ relation of M07 (dashed) steepens the total
profile. The dotted curve, with $M$-$L$ relation of Y08 and $c$-$M$
relation of M07, steepens further but negligibly. The total profile without the
AC (dashed-dotted) is also shown.  {\it Right:} The effect of varying
$\alpha_{\rm LF}$ on the ISD is shown for the total profile. The total profile
without any bias (long dashed) is independent of $\alpha_{\rm LF}$. Changes in
various model parameters have degenerate effects on the expected ISD. The
current uncertainties on the data do not have further constraining power on the
tested model parameters. All the theoretical curves are multiplied by the
P($>\theta_{\rm 3.7\arcsec}$) since the \SA sample consists of lenses with $\theta >
3.7\arcsec$.}
\end{center}
\end{figure*}

O06 assumed the background source population to lie at a redshift of $z_s=2$,
and the source luminosity function, $\Phi(z_s,L)\propto L^{-2.1}$ appropriate
for the radio survey CLASS \citep{myers03,rusin01} \footnote{The luminosity
density for such a steep faint end slope ($\alpha_{\rm LF}<-2$) diverges as
$L\rightarrow 0$ and necessarily requires a cutoff below some value of
$L_{\rm min}$. }. We are able to reproduce the expected ISDs from O06,
given by Eq.~\ref{eq:isdog}, for all the three density distributions
mentioned above. The expected ISDs corresponding to these three density
distributions are shown in Fig.~\ref{fig:isdog}. 

\subsubsection{Observed distribution}
\label{sub:obsisd}
The observed ISD is calculated by logarithmically binning the image separations
of 125 \SA candidates\footnote{Two of the candidates are excluded since their
lensing configurations or the centers of their lens potential were ambiguous.
These candidates have $R_{\rm A}=0.0$\arcsec in Table~\ref{tab:all}.} with
$\theta_{\rm cut}\ge20$~pixels (that is, $>3.7\arcsec$) and an average ranking of 2
and above. The image separation for each lens candidate is taken as twice the
Einstein radius or roughly the arc radius which is the distance between the
candidate lensed image and the center of respective lens galaxy. Let
$\theta-d\theta=\theta_{\rm l}$ and $\theta+d\theta=\theta_{\rm h}$, then the
observed ISD is given by 
\begin{eqnarray}
\bigg|\frac{dP_{\rm lens}}{d\theta}\bigg|_{\rm obs}=\frac{P(>\theta_{\rm h}) - P(>\theta_{\rm l})}{2d\theta} \nonumber \\ 
\label{eq:dpobs}
=\frac { N(>\theta_{\rm h}) - N(>\theta_{\rm l})}{N(>\theta_{\rm cut}) 2d\theta} \, .  
\end{eqnarray}
where the total number of observed lenses is $N(>\theta_{\rm cut})$. 
While comparing their theoretical predictions ($dP/d\theta$) to the observed
ISDs from the CLASS sample, O06 assumed an arbitrary normalization for their
data points. Instead we note that,
\begin{eqnarray}
P_{\rm lens}(>\theta) &=&\frac{N_{\rm lens}(>\theta)}{N_{\rm
lens}(>\theta_{\rm cut})} \nonumber \\ 
&=&\frac{N_{\rm lens}(>\theta)}{N_{\rm src}}\frac{N_{\rm
                      src}}{N_{\rm lens}(>\theta_{\rm cut})}\,, \nonumber \\
&=& \frac{P(>\theta)}{P(>\theta_{\rm cut})}\,,\\
\implies 
\bigg|\frac{dP}{d\theta}\bigg| &=&
\bigg|\frac{dP_{\rm lens}}{d\theta}\bigg|\,P(>\theta_{\rm cut})\,.
\label{eq:isdognm}
\end{eqnarray}

To facilitate a direct comparison of the observed ISD
to the theoretical expectation from O06 (see Eq.~\ref{eq:isdog}), we multiply
Eq.~\ref{eq:dpobs} by $P(>\theta_{\rm cut})$. The quantity $P(>\theta_{\rm
cut})$ is obtained by integrating Eq.~\ref{eq:isdog} from $\theta_{\rm cut}$ to
$\infty$. The \SA data points are shown in magenta in Fig.~\ref{fig:isdog}. The
vertical error-bars are calculated assuming Poisson number statistics and the
horizontal bars show the bin width. 

The \SA data points demonstrate that the average density profile of the halos,
giving rise to the intermediate $\theta$ values ($\sim3\arcsec-12\arcsec$), is best
represented by a combined profile for the main galaxy and the dark matter halo
as opposed to a pure SIS or pure NFW profile. It is clear from the
Fig.~\ref{fig:isdog} that the \SA sample follows a steeper ISD at the
intermediate scales compared to halos with NFW profile and shallower than halos
with only SIS profile.

It is interesting to compare the \SA sample with the SLACS and MACS samples
which span the lower end and higher end of the ISD, respectively. These samples
also have different selection functions compared to the \SA sample. We apply the
same procedure to calculate the observed ISD for SLACS and MACS data points
shown in Fig.~\ref{fig:isdog}. Intriguingly, the SLACS data points lie in the
regime where the SIS density profile just ceases to be dominant.  Although the
SLACS sample appears to be incomplete by a factor of 2 to 3 in the lowest
$\theta$ bin, the ISD of SLACS could be seen as an extrapolation of the ISD from
the \SA sample. As expected, the MACS sample is nearly consistent with either
the NFW or total profile.  Assuming that incompleteness is the only major factor
in the ISD of SLACS, the expected ISD corresponding to the total profile best
matches the SLACS, \SA and MACS samples combined.

We note that we have not accounted for any effects due to purity or
incompleteness of the \SA sample in this paper. We suspect that the completeness
of the sample as a function of the image separation is not severely affected due
to the selection function. We are currently investigating this issue and the
results will be presented in a forthcoming paper.

\subsubsection{Tests with varying models}
Here, we test the effects of varying the different components of the O06 model
and compare against the observed ISD. Since different models have different
values of $P(>\theta_{\rm cut})$, we predict $dP_{\rm lens}/d\theta$ from each
model and compare it with the observed ISD in Fig.~\ref{fig:isdts}. We
test only for image separations spanning the observed $\theta$ values. In the
Fig.~\ref{fig:isdog} and both the panels of Fig.~\ref{fig:isdts}, the solid line
represents the same ISD corresponding to the total two-component profile which
accounts for the contraction of the dark matter.  First, we test the influence of
excluding the AC while computing the total profile.  This has the effect of making
the expected ISD shallower as shown by the dashed-dotted curve in the left panel
of Fig.~\ref{fig:isdts}. Prima facie, the AC model fits the data
better than the model without AC. However, as we show below, there are other
degeneracies in the model which prevent us from ruling out the ``no AC" model
at high significance.

Next, we test the effect of using different $c$-$M$ relations on the ISD. For
example, we use the $c$-$M$ relation by \citet[][henceforth, M07]{maccio07}
instead of that given by \citet{bullock01}. The $c$-$M$ relation of M07 is
roughly 15-20\% lower than that of \citet{bullock01}. The combined profile using
the $c$-$M$ relation of M07 is shown by the dashed curve in the left panel of
Fig.~\ref{fig:isdts}. Within the current statistical limits on the data, both
the $c$-$M$ relations appear plausible, although the data appears to slightly
favor the $c$-$M$ relation of M07. Since the $c$-$M$ relations differ
significantly at small and large image separations, we will need additional
samples of galaxy or cluster-scale lenses to test between the different $c$-$M$
relations.

We also test the effect of using a more recent determination of the $M$-$L$
relation obtained by \citet[][henceforth, Y08]{yang08} from a sample of SDSS
groups along with the $c$-$M$ relation of M07 for the combined total profile.
The $M$-$L$ relation of Y08 differ by $\sim$0.2~dex from that of \citet{vale04}
at the intermediate mass regime which is the regime of interest. This appears
to cause a very negligible change in the predicted ISD.

We try to quantify the effect of varying the slope of the source luminosity
function at the faint-end. We show the effect on the combined profile of O06 and
vary the power law index, $\alpha_{\rm LF}$ of the source luminosity function,
$\Phi(z_s,L)\propto L^{\alpha_{\rm LF}}$. This influences the lens
cross-section via the magnification bias. The solid curve in
Fig.~\ref{fig:isdts} shows the expected ISD using the fiducial value of
$\alpha_{\rm LF}=-2.1$, while the dotted and short dashed curves show
the ISD in the right panel, corresponding to $\alpha_{\rm LF}$ equal to -1.7 and
-1.2, respectively. It is evident from the figure that the observed ISD can be
used to constrain the slope of the luminosity function, if the statistical
error bars could be reduced.

We note that the magnification bias factor in the biased lens cross-section is
calculated assuming that the background sources are point sources such as
quasars. However, the background sources corresponding to the lensed arcs are
mostly galaxies with extended surface brightness and their magnification bias
could be negligible \citep[e.g.,][]{narayan93}. Therefore, we calculate the ISD assuming
no bias, that is, by substituting $B(z_{\rm s})=1$ in Eq.~\ref{eq:siglens}. The
long dashed curve in the right panel of Fig.~\ref{fig:isdts} shows the ISD
without the bias.

The data is consistent with all of the above tested models within the
uncertainties. The various scaling relations from our model, that are tested
here, have degenerate effects on the expected ISD. For example, varying the
$c$-$M$ relation has the same effect as changing the slope of the luminosity
function or excluding AC. However, constraints from independent observations
such as dynamics and strong lensing could be used to determine the $c$-$M$
relation. This will allow us to better constrain the slope of the luminosity
function or make more robust statements about AC.

For all the theoretical calculations, we have assumed $z=2$ for the source
redshift. We checked the effect of adopting different source redshifts,
$z_s=1.5$ and $z_s=3.0$ on the predicted ISD. We found that using a higher
(lower) source redshift causes the predicted $\dx P/\dx \theta$ to be larger
(smaller) by roughly a constant factor at all values of $\theta$. However, the
predicted $\dx P_{\rm lens}/\dx \theta$ is not affected because the
corresponding increase (decrease) in $P(>\theta_{\rm cut})$ almost perfectly
cancels out the increase (decrease) in $\dx P/\dx \theta$. This implies that the
expected ISD would not be drastically different had we accounted for the
distribution of source redshifts instead of assuming a single value for the
source redshift.

\section{SUMMARY}
\label{sec:summ}
We have presented the \SA sample from the completed \CW and \CD covering a combined
unmasked area of $\sim$150~deg$^2$ in the sky. The lens sample is compiled through a
semi-automatic technique consisting of using \AF algorithm, followed by visual
inspection and ranking of the candidates. We briefly described the working of
the \AF \citep{alard06} and the modifications implemented in the newer version
of the algorithm. Although the \AF V2.0 is faster, there is still scope for
improving the algorithm in terms of increasing the purity without compromising
the completeness of the arc detections.

We have compiled a total of 127 candidates in the \SA sample, out of which 38 are
found serendipitously. From the complete sample, 54 are promising lens systems
(ranking of 3 and above). A total of 31 systems are almost certain or confirmed
systems, out of which 27 systems have been followed-up via
techniques such as spectroscopy, high-resolution imaging and/or lens mass
modeling. We found 2 radial arc candidates in the \SA sample and both of them
are located in the system SA100. The second radial arc is clearly identified
from the high resolution \HST \ imaging only. Our sample may have more radial
arcs which could be discovered with high resolution imaging. With statistics of
radial and tangential arcs from a homogeneous and a larger sample of lenses,
interesting constraints could be placed on the slope of inner density profiles
of the dark matter distribution.

We have discovered a total of 12 giant arcs ($l/w\ge8$) in our sample.
Statistics with giant arcs is considered to be a good probe of cluster
properties or cosmological parameters but not an easy one. We have presented the
redshift distributions of the lens galaxies with giant arcs and all arcs in the
\SA sample using the photometric redshifts and found to have mean values at
$z\sim0.6$.  This is somewhat higher than the expected peak at redshift of
$0.3-0.4$ \citep{bartelmann98} but consistent within the uncertainties. Note
that the predictions need to be revised with improved simulations and more
realistic assumptions. We also calculated the angular distribution of giant arcs
which are sensitive to the ellipticity of the halo.  We found an anisotropy in
the orientation of arcs in our sample consistent with that seen in simulated
lens clusters (e.g. see D04).  In addition, the angular distribution of all arcs
from the \SA sample exhibits similar anisotropy. This anisotropy appears to hold
at all arc radii within the current uncertainties albeit needs to be verified
with a sample of confirmed lenses. Thus, we have extended this result to group
scale halos, observationally. It would be interesting to check if lensed arcs in
simulated group scale halos also show a similar anisotropy in their azimuthal
distribution and what constraints could be placed on the baryonic physics
important in the inner few Kpcs which is probed by these lensed arcs.

We followed the formalism of O06 to calculate the expected ISD in order to
compare it with the observed ISD from \SA sample. We first reproduced the
results of O06 and then introduced variations in different scaling relations
used in their model. The \SA sample allowed, for the first time, to probe the
intermediate mass regime corresponding to group-scale halos via the ISD. We
showed that the density profile of the halos are well-reproduced by a combined
profile (NFW and Hernquist) at the group-scales, which is consistent with the
predictions. Given the uncertainties in the data and the degeneracies in the
model, both the models that account for or exclude the role of AC are consistent
with the observed ISD. With the availability of larger statistics of confirmed
lenses and understanding of the sample selection function, the distinction
between models with and without AC would be possible. 

Next, we varied the $c$-$M$ relation, the halo mass-luminosity relation, the
slope of a power-law source luminosity function and the source redshift. We
found that given the current uncertainties in the observed ISD, the $c$-$M$
relations of both \cite{bullock01} and the more recent, \cite{maccio07}, are plausible. The
expected ISD does not vary significantly and fits the data well, if the halo
mass-luminosity relations have an uncertainty of $\sim0.2$~dex. Following O06,
we adopted a power-law index of $\alpha_{\rm LF}=-2.1$ for the background source
luminosity function to account for the magnification bias. We further tested the
effect of varying the $\alpha_{\rm LF}$ on the expected ISD and found to be
consistent with the data. However, within the uncertainties, the data is also
consistent, if no magnification bias is assumed. We did not test the effects of
any evolution of the luminosity function or any other functional form such as a
broken power law. 

We found that varying the model parameters have degenerate effects on the ISD,
for instance, changing the $c$-$M$ relation and changing the slope of the
luminosity function, $\alpha_{\rm LF}$. Therefore, using priors from independent
methods on one of these relations could help in constraining the others via
the observed ISD. Since the background sources are assumed to be at a
redshift of 2, we tested the effect of varying the source redshift. The
expected ISD ($\dx P_{\rm lens}/\dx \theta$) is not affected by choosing
different redshifts (z$_s=1,2,3$) between the range we tested.

As described above, we have used arcs statistics to probe the average density
profiles of group-scale lenses and we have shown the possibility to use arcs
statistics in constraining some scaling relations. However, the models assumed
in our work are simplistic and will need refinement as the lens samples become
larger with upcoming surveys. On the observational side, understanding the
selection effects will also become crucial, if the model parameters need to be
constrained with high accuracy. We hope to address some of these important
issues in future studies.

\acknowledgments

The authors acknowledge support from CNRS and the ANR grant ANR-06-BLAN-0067.
RG acknowledges support from the Centre National des Etudes Spatiales. VM
gratefully acknowledges support from FONDECYT through the grant 1090673. AM
thanks Neal Dalal, Sherry Suyu and Tomas Verdugo for useful suggestions. AM also
appreciates comments from the referee which improved the structure and content
of the paper. The authors recognize and acknowledge the very significant
cultural role and reverence that the summit of Mauna Kea has always had within
the indigenous Hawaiian community. We are most fortunate to have the opportunity
to conduct observations from this mountain. Tests with varying models of ISD
used in this work have been performed on the Joint Fermilab - KICP
Supercomputing Cluster, supported by grants from Fermilab, Kavli Institute for
Cosmological Physics, and the University of Chicago.

{\it Facilities:} \facility{CFHT}, \facility{VLT:Antu}

%


\appendix
\section{A. NUMERICAL DETERMINATION OF LENSING QUANTITIES}

We describe how we numerically determined the relation between the halo mass and
the image separation, the size of radial caustic and the position dependent
magnification for general spherically symmetric mass distributions.
These are used in the calculation of the expected ISD in the two of the models
we tested, namely, the NFW and the total (NFW and Hernquist) profiles.

Let the position of the lens in the plane of the sky be the origin
of the coordinate system. The lensing equation relates the true
angular position, $\vecb{\beta}$, of the background source to the
observed angular position of its image, $\vecb{\theta}$ in the plane
of the sky, such that
\begin{equation}
\vecb{\beta} = \vecb{\theta} - \vecb{\alpha}(\vecb{\theta}) \, ,
\end{equation}
where $\vecb{\alpha}$ is the scaled deflection angle. In case of a
spherically symmetric density distribution for the lens, the
deflection angle $\vecb{\alpha}$ lies along the same direction as
$\theta$, and the lens equation can be written as
\begin{equation}
\vecb{\beta}=\vecb{\theta}-\frac{\alpha(\theta)}{\theta}\vecb{\theta} \, ,
\label{eq:lens}
\end{equation}
where the quantities in normal face are the magnitudes of the
corresponding vectors in bold face. The relation between the deflection
angle $\alpha$ and the magnitude of the image position $\theta$ is given by
\begin{equation}
\alpha(\theta)=\frac{4GM(<D_l\theta)}{c^2D_l\theta} \, ,
\end{equation}
where $D_l$ is the angular diameter distance to the lens and $M(<D_l\theta)$
denotes the projected lens mass within a physical radius of size
$\xi=D_l\theta$, which can be obtained using the following equation
\begin{equation}
M(<\xi) = 2\pi \int_{0}^{\xi} R' \dx R' \int_{0}^{R_{\rm max}} 2 \, 
\rho\left(\sqrt{z^2+R'^2}\right) dz \,,
\label{eq:proj}
\end{equation}
where $R_{\rm max}=\sqrt{R_{\rm vir}^2-\xi^2} $ and $R_{\rm vir}$ is the virial
radius. 

\begin{figure}
\begin{center}
\includegraphics[scale=0.44]{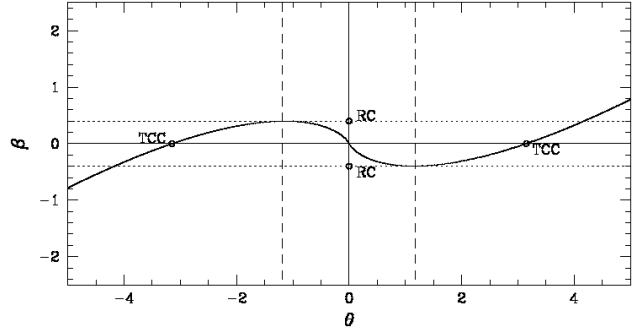}
\caption{ \label{fig:appa}
Source position as a function of image position for spherically symmetric lens
density distribution. The center of the lens potential is at the origin. RC is
the location of the radial caustic and TCC is the location of the tangential
critical curve which corresponds to the Einstein radius.
}
\end{center}
\end{figure}

In Fig.~\ref{fig:appa}, we show the generic shape of how $\beta$ and $\theta$
are related to each other for a spherically symmetric mass distribution which
has a finite density at the center (non-singular). For large values of $\beta$,
$\theta\approx\beta$ corresponding to the weak lensing regime. As the source
approaches the lens in projection, i.e as $\beta \rightarrow 0$, the source
position $\beta$ corresponds to multiple values of $\theta$ referred to as the
strong lensing regime. The first instance of this is when the source is at the
radial caustic (labeled as RC in Fig.~\ref{fig:appa}) and here, the numerical
derivative of $\beta$ with respect to $\theta$ is zero. When $\beta \approx 0$,
$\theta$ corresponds to the Einstein radius ($\theta_E$) which is the location
of the tangential critical curve (see TCC in Fig.~\ref{fig:appa}). The image
separation is approximately equal to $2\theta_E$.

The magnification of the lensed image at $\theta$ is given
by the determinant of
\begin{equation}
\mu(\theta) = \left| \frac{\partial \vecb{\beta}}{\partial
\vecb{\theta}} \right|^{-1} \,.
\end{equation}
In component notation, the required derivative is given by
\begin{eqnarray}
\frac{\partial\beta_i}{\partial \theta_j} &=&
\delta_{ij}\left(1-\frac{\alpha}{\theta}\right) -
\frac{\partial}{\partial \theta}\left( \frac{\alpha}{\theta} - 1
\right)\frac{\partial \theta}{\partial \theta_j}\theta_i \\
&=&
\delta_{ij}\frac{\beta}{\theta}+
\frac{\partial}{\partial \theta}\left( \frac{\beta}{\theta}
\right)\frac{\theta_i\theta_j}{\theta} \\
&=&
\delta_{ij}\frac{\beta}{\theta}+
\left( \frac{\partial \beta}{\partial \theta} - \frac{\beta}{\theta}
\right)\frac{\theta_i\theta_j}{\theta^2} \,.
\end{eqnarray}
The inverse of the determinant of the above equation simplifies to
\begin{equation}
\mu(\theta) = \frac{\theta}{\beta} \frac{\partial \theta}{\partial \beta} \,.
\label{eq:mag}
\end{equation}

To summarize, given a spherically symmetric density distribution
$\rho(r)$, we first obtain the value of $\beta$ for different values
of $\theta$ using equations \ref{eq:lens}-\ref{eq:proj}. We use cubic
splines to approximate the function $\beta(\theta)$. The value of
$\beta$ which corresponds to the local extrema of this function gives
the angular size of the radial caustic. The image separation is given
by two times the value of $\theta$ when $\beta=0$. The mass
within this angular image separation can be calculated using Equation
\ref{eq:proj}. Finally, the magnification as a function of $\theta$
can be obtained using Equation \ref{eq:mag}. The required derivative
is calculated by using the cubic spline fit. The magnification bias
can thus be calculated using Equation \ref{eq:magbias}.

\section{B. MEASUREMENT OF THE PEAK OF THE REDSHIFT DISTRIBUTION}

Consider a lensing galaxy with true redshift $z_t$ for which the photometric
redshift estimate is $z_i\pm\sigma_i$. Let us assume that the true redshifts of
the lens population follow a Gaussian distribution, $G(\bar{z}_t,\sigma_t^2)$, with
the mean and variance given by $\bar{z}_t$ and $\sigma_t^2$, respectively.  The
probability that the estimated redshift of the lens galaxy is equal to $z_i$, is
then given by
\begin{eqnarray}
P(z_i|G)&=&\int P(z_i|z_t) P(z_t|G) dz_t \,,\\
&=& \frac{1}{2\pi} \int \frac{1}{\sigma_i \sigma_t}
{\rm exp} \left[ - \left( \frac{(z_i-z_t)^2}{2\sigma_i^2} +
\frac{(z_t-\bar{z}_t)^2}{2\sigma_t^2} \right)\right] dz_t \,. \\
\end{eqnarray}
Rewriting the factor inside the exponential as a quadratic in $z_t$ and completing the square
yields
\begin{equation}
P(z_i|G)=\frac{1}{2 \pi \sigma_i \sigma_t} 
{\rm exp} \left[ \frac{-(z_i-\bar{z}_t)^2}{ 2(\sigma_i^2+\sigma_t^2) } \right] 
\int {\rm exp} \left[ \frac{-\big[ z_t - \tilde{z}\big]^2}{2} \left( \frac{1}{\sigma_i^2} + \frac{1}{\sigma_t^2}\right)  \right]  dz_t\,,
\end{equation}
where $\tilde{z}=(\sigma_i^2+\sigma_t^2)^{-1} (z_i\sigma_i^{-2} + {\bar
z}_t\sigma_t^{-2})$. Carrying out the integral gives
\begin{eqnarray}
P(z_i|G)&=&\frac{1}{\sqrt{2 \pi (\sigma_i^2 + \sigma_t^2)}} 
{\rm exp} \left[ \frac{-(z_i-\bar{z}_t)^2}{ 2(\sigma_i^2+\sigma_t^2) } \right] \,.
\end{eqnarray}
Since the determinations of the photometric redshifts of lens galaxies are
independent of each other, the combined likelihood for the data, given our
model, can be written as $\mathscr{L}= \prod_{i=1}^{N} P(z_i|G)$. 
The posterior
distribution for our model parameters given the data is then given by Bayes'
theorem
\begin{eqnarray}
P( \bar{z}_t,\sigma_t^2 | z_i,\sigma_i ) \propto \mathscr{L}\,
P(\bar{z}_t,\sigma_t^2)
\end{eqnarray}
We assume uninformative priors on the parameters $\bar{z}_t$ and $\sigma_t^2$
and sample from the above posterior distribution using a Monte Carlo Markov
Chain. We quote the mean of the redshift distribution and the 68 percent
confidence interval on it using the samples from the chain.

\end{document}